\documentclass[15pt,preprint,numberedappendix,appendixfloats,deluxetable]{emulateapj}

\usepackage{graphicx,amsmath,natbib, hyperref}

\slugcomment{{\bf Submitted to {\it Astrophysical Journal} 12 November 2012. Revised 11 February 2013 and 7 March 2013. Accepted 8 March 2013}}

\shorttitle{Reconnaissance of the HR 8799 Exosolar System I}
\shortauthors{Oppenheimer et al.}

\begin{document}

\title{Reconnaissance of the HR 8799 Exosolar System I: Near IR Spectroscopy}

\author{B. R. Oppenheimer\altaffilmark{1}}
\author{C. Baranec\altaffilmark{2}}
\author{C. Beichman\altaffilmark{3, 1}}
\author{D. Brenner\altaffilmark{1}}
\author{R. Burruss\altaffilmark{4}}
\author{E. Cady\altaffilmark{4}}
\author{J. R. Crepp\altaffilmark{5,12}}
\author{R. Dekany\altaffilmark{2}}
\author{R. Fergus\altaffilmark{6}}
\author{D. Hale\altaffilmark{2}}
\author{L. Hillenbrand\altaffilmark{5}}
\author{S. Hinkley\altaffilmark{5}}
\author{David W. Hogg\altaffilmark{7}}
\author{D. King\altaffilmark{8}}
\author{E. R. Ligon\altaffilmark{4}}
\author{T. Lockhart\altaffilmark{4}}
\author{R. Nilsson\altaffilmark{1}}
\author{I. R. Parry\altaffilmark{8, 1}}
\author{L. Pueyo\altaffilmark{9}}
\author{E. Rice\altaffilmark{10, 1}}
\author{J. E. Roberts\altaffilmark{4}}
\author{L. C. Roberts, Jr.\altaffilmark{4}}
\author{M. Shao\altaffilmark{4}}
\author{A. Sivaramakrishnan\altaffilmark{11, 1}}
\author{R. Soummer\altaffilmark{11, 1}}
\author{T. Truong\altaffilmark{4}}
\author{G. Vasisht\altaffilmark{4}}
\author{A. Veicht\altaffilmark{1}}
\author{F. Vescelus\altaffilmark{4}}
\author{J. K. Wallace\altaffilmark{4}}
\author{C. Zhai\altaffilmark{4}}
\author{N. Zimmerman\altaffilmark{1,13}}

\altaffiltext{1}{Astrophysics Department, American Museum of Natural History, Central Park West at 79th Street, New York, NY 10024 USA; e-mail inquiries should be directed to bro@amnh.org.  A description of the contributions of each author can be found at \url{http://www.amnh.org/project1640}}
\altaffiltext{2}{Caltech Optical Observatories, California Institute of Technology, Pasadena, CA 91125 USA}\altaffiltext{3}{NASA Exoplanet Science Institute, California Institute of Technology, Pasadena, CA 91125 USA}
\altaffiltext{4}{Jet Propulsion Laboratory, California Institute of Technology, 4800 Oak Grove Dr., Pasadena CA 91109 USA}
\altaffiltext{5}{Department of Astronomy, California Institute of Technology, 1200 E. California Blvd, MC 249-17, Pasadena, CA 91125 USA}

\altaffiltext{6}{Department of Computer Science, Courant Institute of Mathematical Sciences, New York University, 715 Broadway, New York, NY 10003 USA}
\altaffiltext{7}{Center for Cosmology and Particle Physics, Department of Physics, New York University, 4 Washington Place, New York, NY 10003 USA}
\altaffiltext{8}{Institute of Astronomy, Cambridge University, Madingley Road, Cambridge CB3 0HA United Kingdom}
\altaffiltext{9}{Department of Physics and Astronomy, The Johns Hopkins University, Baltimore, MD 21218 USA}
\altaffiltext{10}{Department of Engineering Science \& Physics, College of Staten Island, 2800 Victory Blvd., Staten Island, NY 10314 USA}
\altaffiltext{11}{Space Telescope Science Institute, 3700 San Martin Drive, Baltimore, MD 21218 USA}
\altaffiltext{12}{Now at Notre Dame University, Indiana USA}
\altaffiltext{13}{Now at Max Planck Institute for Astronomy, Heidelberg, Germany}

\begin{abstract}
We obtained spectra, in the wavelength range $\lambda = 995 - 1769$ nm, of all four known planets orbiting the star HR 8799.  Using the suite of instrumentation known as Project 1640 on the Palomar 5-m Hale Telescope, we acquired data at two epochs.  This allowed for multiple imaging detections of the companions and multiple extractions of low-resolution ($R \sim 35$) spectra.  Data reduction employed two different methods of speckle suppression and spectrum extraction, both yielding results that agree.  The spectra do not directly correspond to those of any known objects, although similarities with L and T-dwarfs are present, as well as some characteristics similar to planets such as Saturn.  We tentatively identify the presence of CH$_4$ along with NH$_3$ and/or C$_2$H$_2$, and possibly CO$_2$ or HCN in varying amounts in each component of the system.  Other studies suggested red colors for these faint companions, and our data confirm those observations.  Cloudy models, based on previous photometric observations, may provide the best explanation for the new data presented here.  Notable in our data is that these presumably co-eval objects of similar luminosity have significantly different spectra; the diversity of planets may be greater than previously thought.  The techniques and methods employed in this paper represent a new capability to observe and rapidly characterize exoplanetary systems in a routine manner over a broad range of planet masses and separations. These are the first simultaneous spectroscopic observations of multiple planets in a planetary system other than our own.
\end{abstract}


\section{Motivation and Properties of HR 8799}\label{intro}
The star HR 8799 (HD 218396; V432 Pegasi), an A5V, $\gamma$ Dor-type variable star \citep{1995IBVS.4170....1R,1998A&A...337..790A, gk99} at 39.4pc \citep{v07}, has been the target of many observations in the past few years due to the discovery of several faint companions in orbit about it \citep{mmb08,mzk10}.  The nature of these sources---whether they are planets or brown dwarfs \citep[e.g][]{mrs10}---has been subject to significant debate.  Various observations have attempted to constrain the age of the system (and thus the masses of the companions based on theoretical cooling models); to obtain as many uncontaminated photometric measurements of the companions as possible; and to determine their orbits. Here we refer to them as planets while acknowledging that some researchers still have doubts about the use of this label.



\begin{deluxetable*}{clcccccc}
\tabletypesize{\scriptsize}
\tablecaption{Observations}
\tablewidth{0pt}
\tablehead{ \colhead{Date}  & \colhead{Julian} & \colhead{n$_{\rm exp}$ x $t_{\rm exp}$} & \colhead{Estimated} & \colhead{Estimated} & \colhead{CAL} & \colhead{rms-WFE } & \colhead{Planets}\\ \colhead{(UT)} & \colhead{Date (start)} & \colhead{(s)} & \colhead{seeing (\arcsec)\tablenotemark{a}} & \colhead{Strehl\tablenotemark{b}} & \colhead{Mode\tablenotemark{c}} & \colhead{(nm)\tablenotemark{d}} & \colhead{Detected}}
\startdata                                     
14 June 2012 & 2456092.972859 & 5 $\times$ 549.91 & 1.4 & 65\% & $\Phi$ only  & 5          & bcde\\  \vspace{1mm}\\
15 June 2012 & 2456093.971030 & 5 $\times$ 366.61 & 1.1 & 67\% & $\Phi$ only  & 4  & bcde \\ \vspace{1mm}\\
5 October 2012 & 2456205.706470 & 18 $\times$ 549.91 & 1.8 & 44\% & $\Phi$ and $\overrightarrow{E}$ & 14 & c
\enddata
\tablenotetext{a}{Seeing was estimated using short images with no AO correction and is quoted as the FWHM of the PSF as measured at $\lambda = 1653 \pm 12.5$ nm.}
\tablenotetext{b}{Strehl was estimated at $\lambda = 1653 \pm 12.5$ nm using unocculted, AO-corrected images of the primary star taken just before coronagraphic occultation and long integrations and does not include the effects of CAL on the mid-spatial frequencies in the image.}
\tablenotetext{c}{``$\Phi$ only'' means phase correction only, while ``$\overrightarrow{E}$'' means full electric field conjugation was applied for optical speckle control.} 
\tablenotetext{d}{rms wave front error in spatial frequencies from 5 to 32$\lambda/D$ as measured by CAL, without seeing.}
\label{observelog}
\end{deluxetable*}

In order to understand the physics and chemistry of these objects, spectroscopy is necessary.  Herein we present the results of high-contrast, direct, spectroscopic imaging observations of this exosolar system after a short review of its properties, as observed to-date.  A companion paper (Paper II; \citealt{paperII13}) reports the astrometric observations of the system, and a third paper (Veicht et al.~2013) examines the dynamical stability of the system.  

For reference, detailed reviews of high-contrast, coronagraphic observations can be found in \cite{2011exop.book..111T}, \cite{am10} and \cite{oh09}.

\subsection{Variability and Rotational Velocity}
The classification of HR~8799 as a $\gamma$ Dor variable \citep{1999PASP..111..840K} was secure by 1997. Periodicity of the variations from {\it Hipparcos} data \citep{1998A&A...337..790A} and a long photometric campaign revealed multiple periodic signals with frequencies ranging from 0.2479 to 1.9791 days with amplitudes up to 1.5\% and a range of $\pm 4\%$ in the longest mode \citep{1999MNRAS.303..275Z}.   The derived stellar properties from this study are $L = 5.0 L_\sun$, $T_{\rm eff} = 7230$K and $R = 1.44 R_\sun$ \citep{2012ApJ...761...57B}. 

Large time-series data sets of photometry and spectra have been taken on HR~8799 and a number of asteroseismology analyses have been conducted, first to constrain the age, unfortunately with only weak constraints, \citep{2010MNRAS.405L..81M} and later to constrain the inclination angle of the star's rotation axis ($i > 40^{\circ}$) \citep{rks09, 2011ApJ...728L..20W}.  The rotational velocity of the star is $ v \sin i = 37.5$ to $49$ km s$^{-1}$, based on a variety of different measurements \citep{2012A&A...537A.120Z,1998MNRAS.294L..35K}.

\subsection{Age and Planet Masses}
Though still under debate, several indicators suggest that the star is younger than 100 Myr \citep{mzk10,dla10, zrs11,Mal13,2010MNRAS.405L..81M}.  One estimate of this system's age comes from two independent  studies \citep{dla10,zrs11}, each of which used galactic space motions and spectroscopic age indicators to conclude that HR 8799 has a very high probability ($\sim$98\%) of being a member of the 30 Myr Columba Association.  In contrast, \citet{hrk10} convincingly demonstrate that this association is poorly supported and that the age of HR~8799 cannot reliably be associated with Columba. Earlier works argued for a system age younger than 50-60 Myr \citep{mmb08,rks09}.  While significantly different from the 1 Gyr estimate put forth by \cite{2010MNRAS.405L..81M} using asteroseismology data, the \citet{2010MNRAS.405L..81M} work does not explicitly exclude such a young age.

The age of the system is important because it constrains the masses of the planets when used in conjunction with theoretical cooling models and the measured luminosity of the objects.  It is interesting to note, as discussed in \citet{she12}, that some of the strongest constraints on the age of the HR 8799 system come from the dynamical simulations of the four planets, combined with their broadband photometry. As \citet{gm09} discuss, only a small fraction of three-body systems resembling HR 8799 can remain stable longer than $\sim$100 Myr.  Further, the stability of a four planet system for $\sim$10-100 Myr is consistent with lower masses for the companions \citep{gm09,rks09, fm10, mrs10}. Specifically, the study presented in \citet{gm09} finds only very limited dynamical configurations that foster long term stability: such stability will occur if the planets exhibit low-order, two- or three-body mean motion resonances.

\begin{figure*}[htb]
\center
\resizebox{1.0\hsize}{!}{\includegraphics[angle=0.]{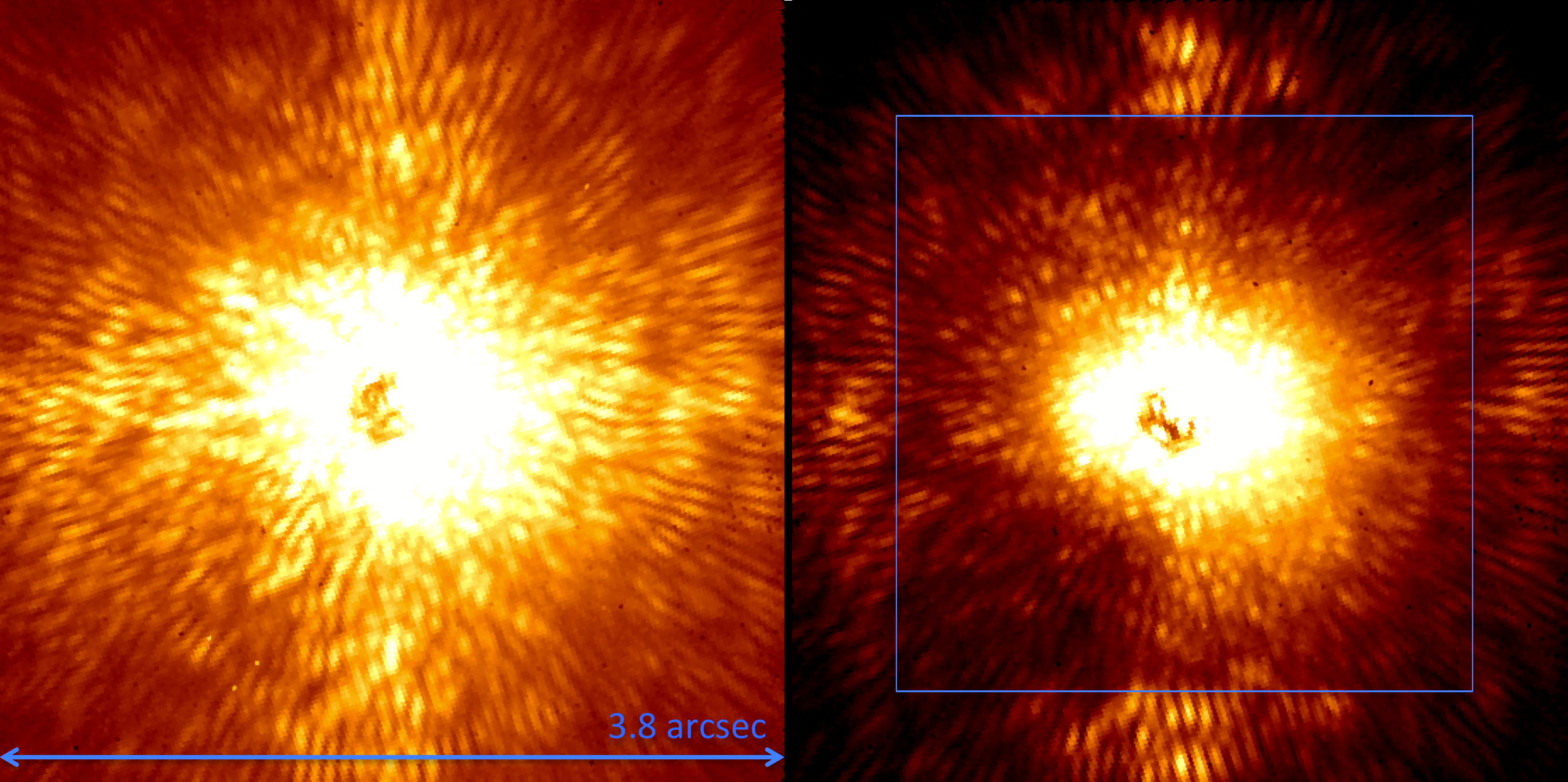}}
  \caption{Images demonstrating the optical suppression of speckles through wave front phase error sensing and control.  Left: coronagraphic image of HR~8799 from 15 June 2012 without the CAL system in operation.  Right: the CAL system iterated 5 times with phase conjugation only (``$\Phi$-only'') to suppress speckles within the 32$\lambda/D$ distance from the star (within the blue square).  In the image to the right, the control region of PALM-3000 (32$\lambda/D$) is clearly apparent as the darker square area centered on the star and extending to the edge of the control region (the blue square), where the speckles brighten again.  Images are both 549-s long and are single slices at $\lambda = 1224$ nm. Color scale is logarithmic and identical in both images. North is up and East to the left, with an image width of 3$\farcs$8.  The planets, whose locations are shown in Fig.~\ref{specklesuppimages}, have brightnesses of 3.2, 3.3, 2.9 and 3.7\% of the mean speckle brightness in the vicinity of each planet, while the CAL system is operating, for b through e respectively.  Thus, advanced image processing techniques are required to study them.}
  \label{rawimage} 
\end{figure*}

These conclusions are consistent with the work presented by \citet{fm10}, which find a longer system lifetime if the two innermost planets maintain a 2:1 commensurability. Further, the persistence of this 2:1 resonance to the current epoch would put a specific constraint on the masses to be less than 10 M$_{\rm Jup}$. Taking this one step further, \citet{fm10} point out that these dynamical constraints serve as early tests of the so called ``hot start'' models \citep{bcb03}. Specfically, if all three planets exhibit low-order commensurabilities with each other, the hot start models are easily consistent with the measured luminosities and inferred masses. If, on the other hand, only the two innermost planets have a 2:1 commensurability, the hot start models are only barely applicable. These studies point to the promise of using high contrast imaging to place empirical constraints on evolutionary models for the brightness of substellar objects \citep[e.g.][]{cjf12}. Nonetheless, the published absolute $H$-band magnitudes, M$_{\rm H}$ $\sim$ 14 to 15 \citep{mmb08}, are consistent with standard cooling evolutionary models  for 30 Myr \citep{bmh97}.

Further, the degree of excess infrared emission measured at 24 $\mu$m \citep[][]{srs09} is consistent with an age significantly younger than 1 Gyr. Observations of A-stars with circumstellar material plus well-constrained ages show a marked decline with age in the strength of the 24 $\mu$m infrared excess \citep[e.g.][]{rss05}. This empirical result can be well-modelled by dynamical simulations \citep[e.g.][]{wss07}, illustrating the clear decay of this hot dust emission due to the shorter dynamical time scales associated with the innermost portions of these systems.  

Based on the ages of 30, 60 and 430 Myr the four companions have masses in the following ranges:  7, 13 and 34 M$_{\rm Jup}$ for c, d and e; and 5, 14 and 23 M$_{\rm Jup}$ for b using the cooling models of \cite{bcb03}.  An important caveat exists regarding the age of the system, as discussed in \cite{2010MNRAS.405L..81M}.  If the inclination angle of the star is near 50$^{\circ}$ and the mass of the star is 1.45 M$_\sun$, the age range from asteroseismology analysis is 26 to 430 Myr, with the larger ages pushing the masses of all components into the ``brown dwarf'' mass range.  As \citet{mrs10} point out, the inclination of the star is probably smaller than the value initially assumed.  Nonetheless, spectroscopy can also be used to constrain mass, or more accurately, $\log(g)$, as described in, for example, \cite{2012ApJ...754..135M} and \cite{2011ApJ...737...34M}. 

\subsection{Previous Spectroscopic Measurements}\label{prevspec}

A number of studies have accomplished spectroscopy of the b component in the $L$, $K$ and $H$ bands \citep{jbg10,bld10,bmk11}.  In addition, preliminary observations of the c component in K band were presented by \cite{2013AAS...22112603K}.  In \cite{jbg10}, a noisy $L$-band spectrum of b, ultimately sub-sampled to very low resolution, suggested that either improvements in the treatment of dust are required to model the object properly, or significant non-equilibrium chemistry is present in b's atmosphere.  \cite{bld10} obtained a $K$-band spectrum of b, and, combined with the \cite{bmk11} data in $H$ and $K$, a number of modeling papers, notably \cite{2012ApJ...754..135M}, suggest that complex cloud structure, high metallicity and/or non-equilibrium chemistry are needed to understand this object.   In any case, b shows little or no methane absorption, possibly some CO and the shape of the $H$-band peak suggested low-gravity \citep{bmk11}.  Unfortunately at the time of writing, the c component spectrum \citep{2013AAS...22112603K} and its analysis were not available in any publication.

These previously published measurements are all of the outer most planet in the system.  However, this currently unique system has four directly detected planets.  Thus it is ideal for the study of multiple planets around a single single star to investigate the diversity of planets.  This is the purpose of this paper.

\subsection{Orbital Characteristics}
HR 8799 has become a benchmark target of observation in the field of direct exoplanet imaging to demonstrate whether a given project or technique is capable of finding even fainter objects, relative to the star, than the four planets orbiting this star.  As a result numerous images of the system have been obtained mainly in the near IR through $L$- and $M$-bands
\citep{hrk10, 2013A&A...549A..52E,cbi11,hci11,2012ApJ...755L..34C,gmm11,jbg10,lmd09,mmb08,mzk10,smb10,she12,Sou11,2012ApJ...755L..34C}.  The conclusion of this work from various different instruments and telescopes indicates that the three outer planets are roughly coplanar,  orbiting at an inclination angle of $i \sim 27.3$ to $31.4^{\circ}$, with eccentricities below 0.1 and semi-major axes of 68, 42 and 27 AU for b, c, and d respectively \citep[see esp.][]{Sou11,2012ApJ...755L..34C}.  Orbital motion of e is poorly constrained at present, and only two groups attempted dynamical studies including all four planets \citep{2013A&A...549A..52E,mzk10}.  

The astrometry of this system is complicated by the background speckle field and, due to the extensive nature of the data reduction, we present that analysis in a second paper \citep[Paper II;][in prep.]{paperII13}.  However, HR 8799 will likely be the first exoplanetary system with a complete characterization including dynamical masses from astrometry through atmospheric analysis via spectroscopy, part of the purpose of combining imaging with spectroscopy in a single instrument such as Project 1640.  We defer further discussion of the orbital characteristics to Paper II.


\begin{figure*}[]
\center
\resizebox{1.0\hsize}{!}{\includegraphics[angle=0.]{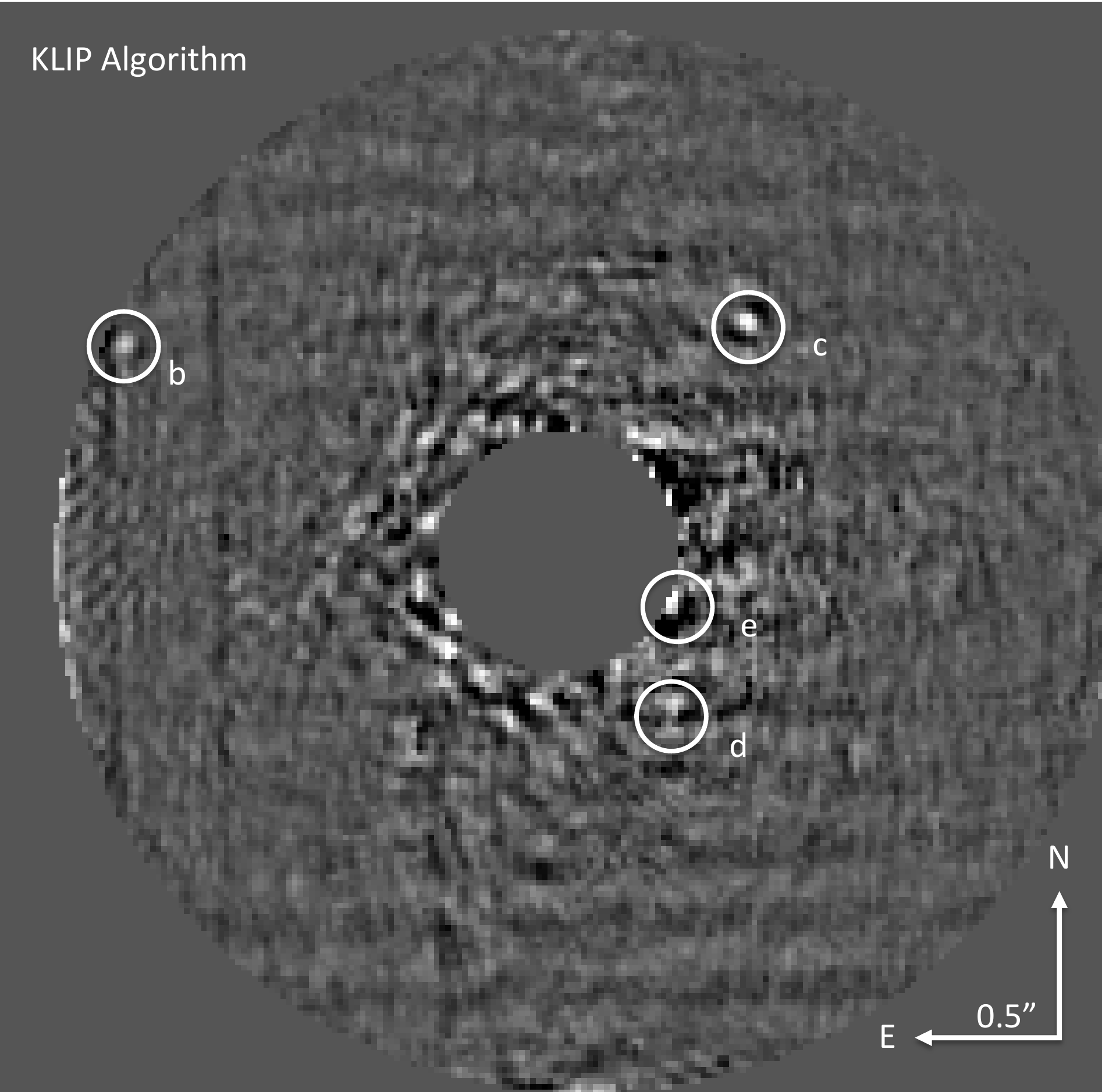} \includegraphics[angle=0.]{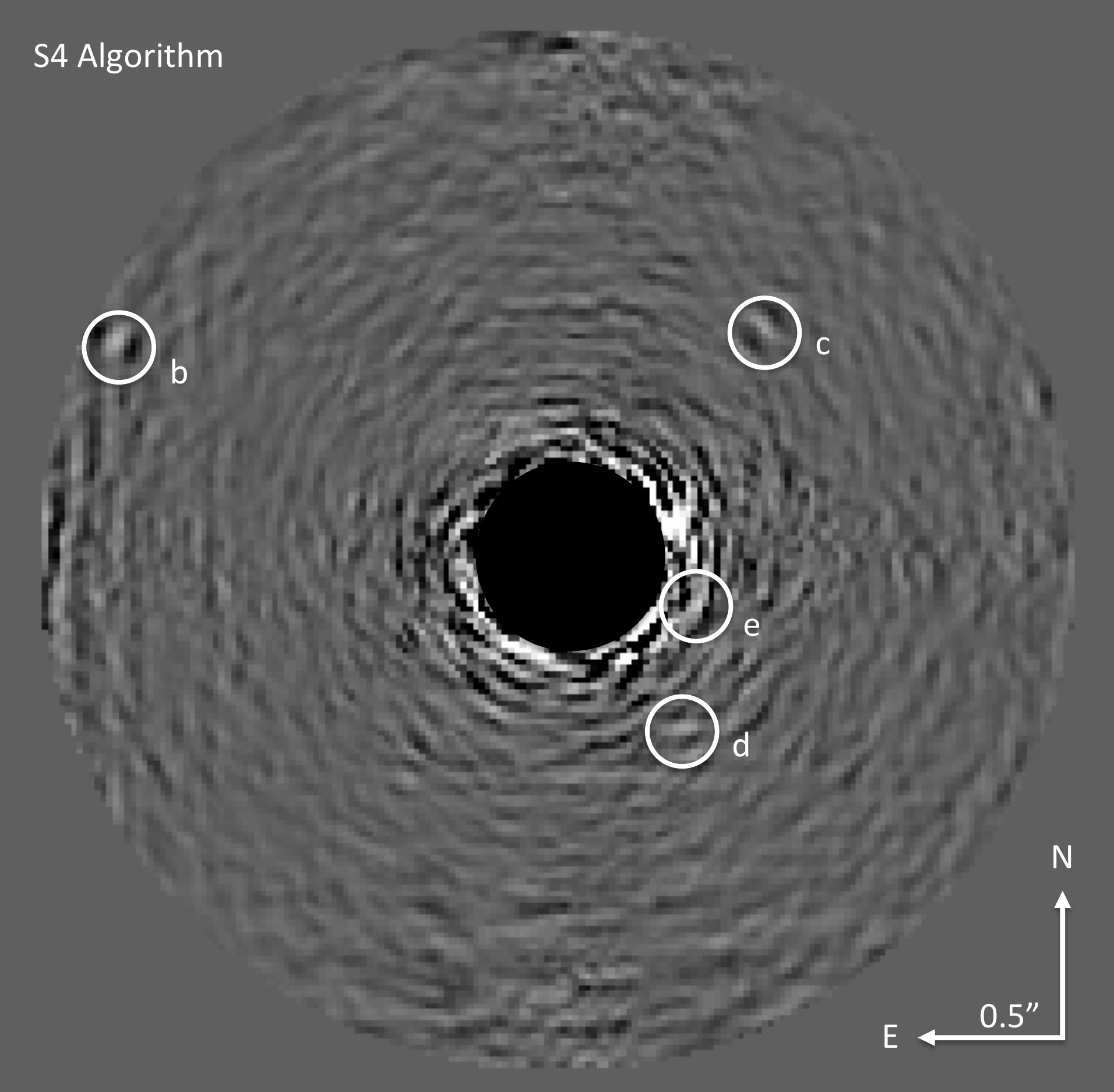}}
  \caption{Speckle suppressed images of the HR 8799 system from 14, 15 June 2012. Left: Speckle suppression by the KLIP algorithm for June 2012 (summed over both dates) in $H$-band.  Right: same as left but speckle suppression was achieved by the S4 algorithm. These images both represent a total of 4582.61-s of exposure time.  In October 2012 only the c component was detected due to poor seeing conditions and an incorrect field-of-view setting that vignetted the position of the b component (see text). Because of differences in the two algorithms, the blocked out areas in the center of the images are different.}
  \label{specklesuppimages} 
\end{figure*}

\subsection{Circumstellar Material}
The HR 8799 system was first identified by \citet{zs04} as a source with a prominent IRAS infrared excess, indicating the presence of cool dust.  Shortly thereafter, the analysis of Spitzer spectroscopic data presented in \citet{csb06} indicated that the debris structure in this system was organized into two spatially distinct components. After the initial discovery of the planets in the system, \citet{srs09} presented the analysis of much deeper Spitzer spectroscopy as well as photometry at 24 and 70 $\mu$m.   In addition to constraining the spatial extent of the dust belts, this work identified an extensive outer halo of dust particles at hundreds of AUs, and these larger regions were later probed through observations presented in \citet{2011ApJ...740...38H}, and \citet{pbk11} at 350 and 880 $\mu$m, respectivley. These observations were used to resolve the outer debris structure surrounding the planets, solidifying the framework presented in \citet{srs09}.  A coherent picture of the overall architecture of the system near the time of the planet discovery, including the third outer component to the debris disk, was synthesized in \citet{rks09}

The inner structure of the HR 8799 debris disk is shaped through dynamical interactions with the four planets.  Specifically, the radial gap in the debris structure from 15-90 AU is clearly consistent with clearing caused by the four known planets. However, the source of clearing interior to 6 AU in the inner warm belt remains a mystery, with only \citet{hci11} placing firm constraints in this region.  If a planetary mass companion is responsible for the clearing, \citet{hci11} put a firm upper limit of $\sim$11 M$_{\rm Jup}$ on its mass, ruling out the possibility of a companion star or brown dwarf between 0.8 and 10 AU.  Spectroscopy in the published literature has not revealed any evidence of binarity either.

\section{Instrumentation: Project 1640}\label{instrument}
Project 1640 is described in \citet{obb12}, \citet{hoz11}, \citet{hob08} and in detail at the level of circuit diagrams, cryogenics, control software, interfaces and opto-mechanical design in \citet{h09}.  The latest system performance metrics are given in \cite{obb12}, including on-sky contrast measurements.  These are described in relation to other projects in high-contrast imaging in \cite{2012SPIE.8442E..04M}, in particular, their Fig.~1.  In summary, the system is capable of producing images with a speckle floor at roughly 10$^{-5}$ at 1\arcsec~separation from a bright star (or 10$^{-7}$ in the lab).  This is achieved through the coordinated operation of four optical instruments: a dual deformable mirror, adaptive optics (AO) system with 3629 actively-controlled actuators, called PALM-3000 \citep[][and 2013, in preparation]{dbr07,dbb06}; an apodized pupil, Lyot coronagraph \citep[APLC;][]{spf09,sl05,s05,saf03eas,saf03}, the design details of which are given in \citep{h09}; a Mach-Zehnder interferometer that senses and calibrates, through feedback to PALM-3000, residual path-length and amplitude errors in the stellar wave front at the coronagraphic occulting spot for optimal diffractive rejection of the primary star's light \citep[CAL;][]{Gautam12,cheng12}; and an integral field spectrograph that takes 32 simultaneous images with a field of view of $3\farcs8 \times 3\farcs8$ spanning the range $\lambda = 995 - 1769$ nm with a bandwidth of $\Delta\lambda = 24.9$ nm per image \citep[IFS, Fig.~\ref{rawimage};][]{hoz11,hob08,h09,obb12}.  Aside from technical advances in high-contrast imaging, numerous results from the project include, among others, the discovery and astrometric and spectroscopic characterization of the Alcor AB system \citep{zoh10}, the $\alpha$ Ophiucus system \citep{hmo11}, the $\zeta$  Virginis companion \citep{hob10}, and comprehensive spectral studies of the companion of FU Orionis \citep{2012ApJ...757...57P} and Z CMa \citep{2013ApJ...763L...9H}.

Raw science data generated by Project 1640 are in the form of 2040 $\times$ 2040 pixel images containing 37146 closely packed spectra roughly 30.4 $\times $ 3.2 pixels in extent.  These images are processed into data cubes with dimensions R.A., $\delta$ and $\lambda$, as described in \cite{zbo11}.  Cross talk between adjacent spatial pixels and across wavelengths is less than 0.4\% flux contamination on average with a maximum of 1.4\% in the water band between $J$  and $H$-band, as confirmed using monochromatic sources and testing of the cube extraction software \citep[see][]{obb12}.  This level of cross talk does not introduce systematic errors in spectral extractions or image properties that would affect relative photometry at a level higher than 2\% to be conservative.

\begin{figure}[]
\center
\resizebox{1.0\hsize}{!}{\includegraphics[angle=0.]{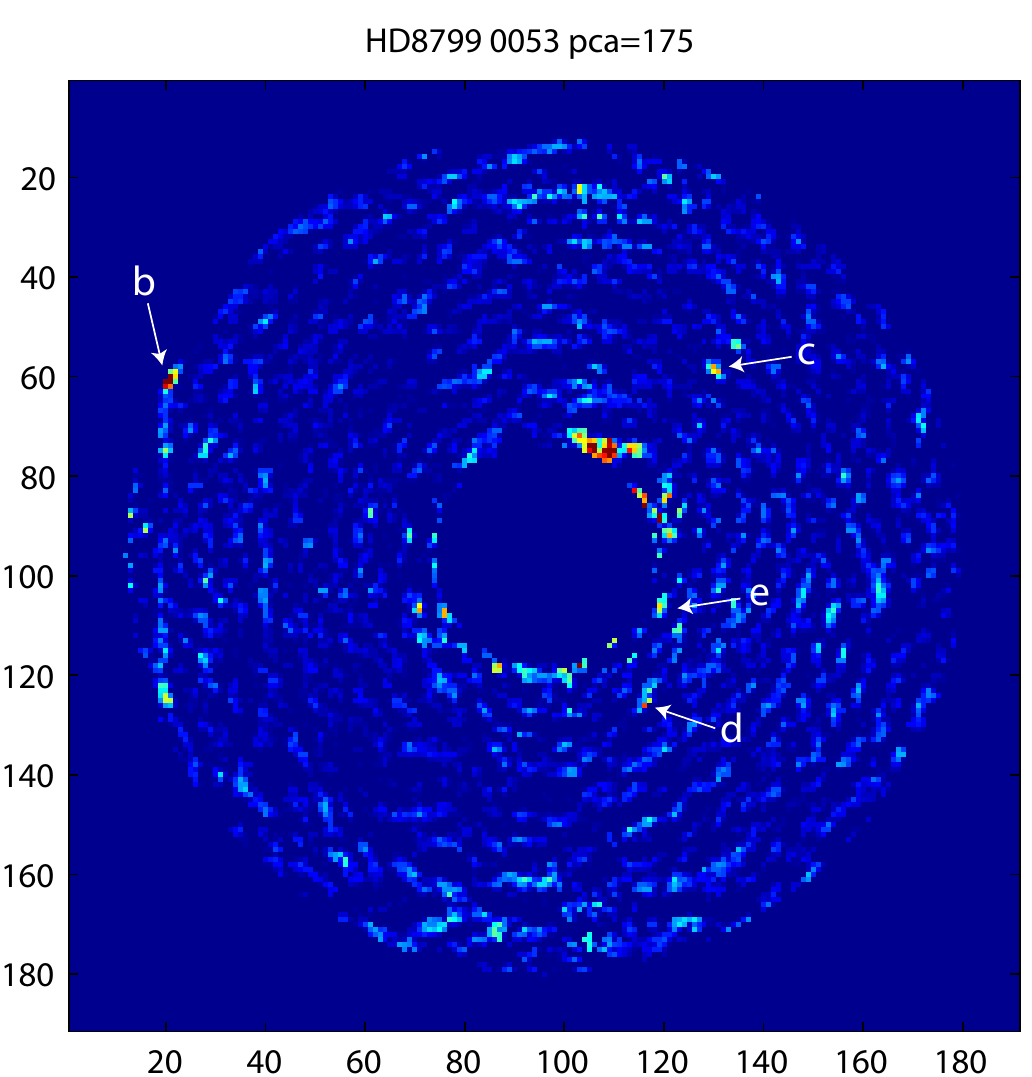} }
  \caption{S4 detection map for June 2012.  Orientation is N up and E left.  Coordinates are in pixels and the detection map is a visualization of the correlation between a companion PSF, obtained experimentally, and the residual image (Fig.~\ref{specklesuppimages} right).  Companions are found with higher fidelity using the detection map than the residual image alone, because it rejects many apparent false positives in the residual image.}
  \label{detmap} 
\end{figure}

Because the planets we observed have relative brightnesses of 3.2, 3.3, 2.9 and 3.7\% of the mean speckle brightness in the vicinity of each planet, for b through e respectively, and the speckles vary spatially and temporally, advanced image processing algorithms are required in addition to the complex suite of hardware and software described above (Figs.~\ref{specklesuppimages} and \ref{detmap}).  

\section{Observations and Data Reduction}
Target acquisition for Project 1640 involves placing the star within the field of of view with telescope pointing, optimizing the AO correction, obtaining images of the star when it is not occulted by the coronagraphic spot (called Core images---acquired only if the star does not saturate the detector in short exposures), acquiring the star behind the coronagraphic spot (using a fine guidance sensor that centers the star on the spot), engaging the CAL system to dim quasi-static speckles \citep[e.g.][]{hos07} and then acquiring deep exposures sensitive enough to detect the speckle floor.  Core images are used for relative photometry and spectral calibration.  If Core images saturate, a fainter star from the IRTF spectral library \citep{rcv09} is used for these calibrations.  In the case of HR 8799, the Core image does not saturate in short (1.45-s) exposures.

The CAL system senses the full electric field of the incident light and can operate in two modes by feeding back different control signals to PALM-3000: (1) suppressing speckles only due to wave front phase errors ($\Phi$ in Table \ref{observelog}), which generates a dark square region centered on the primary star with an angular size of 32$\lambda/D$ extending from the star in all four cardinal directions \citep[Fig.~\ref{rawimage};][]{obb12}, where $D$ is the telescope diameter as modified by any pupil-plane stops in the optical system; or (2) suppressing speckles due to both phase and amplitude errors in the wave front to produce a much darker region in only half of the field of view, extending from the star to 32$\lambda/D$ but only in three of the cardinal directions \citep[NSEW;][]{Gautam12,cheng12}.  While the second mode works, the observations used in this paper were primarily with phase-only ($\Phi$) control, with about half of the October 2012 observations using electric field conjugation.

Between long exposures, short exposures ($t_{\rm int} = 30$-s) for the purposes of precision astrometry \citep[$\sigma \simeq 2$ mas;][]{zoh10} were obtained in which a 150 nm amplitude sine wave was induced on the high-order deformable mirror in two orthogonal directions with 22 cycles each over the telescope pupil.  These create four spots, at a separation of $22\lambda/D$, that permit precise determination of the position of the star behind the coronagraphic mask.  The details and use of these spots have been described elsewhere and will be dealt with in more detail in Paper II \citep{dho06,mlm06,so06,zoh10}.

Observations of HR~8799 on 14 and 15 June 2012 comprised a total of 46 and 31 minutes of exposure time, while 165 minutes of integration time were obtained on 5 October 2012, respectively.  Details of the observations and conditions are provided in Table \ref{observelog}.  Astrometry images (as described above) were not obtained in June 2012 but were taken in October 2012.  Examples of one wavelength slice of two different cubes are shown in Fig.~\ref{rawimage} before any speckle suppression post-processing.  The right image in Fig.~\ref{rawimage} shows the effect of the CAL system, in $\Phi$-only mode, on the stellar point spread function (PSF), creating a darker square centered on the star, within the AO control region (32 $\lambda$/D).

Our experience with the PALM-3000 AO system shows that Project 1640 can only achieve its full sensitivity when the natural seeing is below about $1\farcs7$, which was the case during the June 2012 observations.  In October 2012 however, conditions were considerably worse.  In addition, due to a slight misalignment between the IFU and coronagraph, the b component was outside the field of view in October 2012.  As a result, the October 2012 data  reveal only the c component of the HR~8799 system.  It is important to note that the inner planets were not seen in October, despite the significantly longer exposure time, because of the larger residual wave front error (WFE) of 14 nm rms as opposed to the 4 and 5 nm achieved in June.  Speckle brightness is a very strong function of rms-WFE when it is well below the 50 to 150 nm achieved by most AO systems at these wavelengths.  As such, the speckles were more than one hundred times brighter than the planets in October, making them invisible, even to the advanced speckle suppression techniques described below.

\begin{deluxetable*}{ccccc}
\tabletypesize{\scriptsize}
\tablecaption{Absolute Photometry of the HR~8799 System}
\tablewidth{0pt}
\tablehead{\colhead{} & \colhead{HR~8799 b} & \colhead{c} & \colhead{d} & \colhead{e}}
\startdata
\\
Project 1640 M$_{\rm J}$ &16.48 $\pm$ 0.18 & 15.36 $\pm$ 0.21 &$ >15.5 $ & $> 13.2$\\
Published\tablenotemark{a} M$_{\rm J}$ & 16.30 $\pm$ 0.16 & 14.65 $\pm$ 0.17 & 15.26 $\pm$ 0.44 & No reported value\\\\
 \hline\\
Project 1640 M$_{\rm H}$ &15.11 $\pm$ 0.12 & 14.20 $\pm$ 0.15 & 13.69 $\pm$ 0.19 & 13.30 $\pm$ 0.27 \\
Published$^a$ M$_{\rm H}$ & 14.87 $\pm$ 0.17 & 13.93 $\pm$ 0.17 & 13.86 $\pm$ 0.22 &  13.53 $\pm$ 0.44\\
 \enddata
 \tablenotetext{a}{Published values are from \citet{mmb08} and \citet{mzk10} and the primary has M$_{\rm H}$ = 2.30 \citep{csv03}.\\}
 \label{photom}
 \end{deluxetable*}

\section{Speckle Suppression}
Raw images (Fig.~\ref{rawimage}) from the extracted cubes do not directly reveal the companions of HR~8799.  The speckles in the outer parts of the PSF have to be removed in an effective manner or modeled precisely.  We have used two different techniques, each of which employs a new application of principal component analysis in astronomy.  The first method, called KLIP \citep{2012ApJ...755L..28S}, operates to suppress speckles within a given cube, removing the speckles from each slice individually.  Details on this algorithm are given in Appendix \ref{klipsect}.  The second technique, S4  \citep{s4}, uses the full depth of the diversity of measurements in the data, including modeling the changes in the speckles in time, from cube to cube.  See Appendix \ref{s4sect} for more details.

Fig.~\ref{specklesuppimages} presents images of the system after speckle suppression with KLIP (left) and S4 (right).  Note that all four companions are visible, though the S4 algorithm accomplishes higher signal-to-noise ratio (SNR).  For each component the maximum SNR at any wavelength is 15.4, 29.1, 9.7 and 8.9-$\sigma$ for b, c, d and e respectively (Table~\ref{sigma}).

The S4 algorithm uses the residual image to identify possible companions in a given set of data automatically.  This is done by cross-correlating the residual images with a model companion PSF, which is actually real data acquired with the laboratory white-light source over a grid of 25 positions on the field of view.  These 25 PSFs are used to account for any variation in PSF shape and size over the range of positions and wavelengths in the cubes.  The algorithm assumes a ``white'' spectrum and generates, via the PSF cross-correlation, what we call the ``detection map.''  This map for the June 2012 observations is shown in Fig.~\ref{detmap} (for 200 components in the PCA model; see Appendix \ref{s4sect}).  In this image, positive correlations can be as sharp as a single pixel, and one can see that all four planets are clearly detected, with higher SNR than in the residual image.  This is partly why S4 is such an effective speckle suppressor. 

Although no {\it a priori} information on the locations of the four components was used in the detection algorithms, the e component would not have been discovered with these techniques, at least on simple examination of the detection map and the residual image produced by S4.    However, there is {\it a priori} information on this system from other observations, so we were able to retrieve the e component.  We note that an extension to the algorithm examines the extracted spectra of all peaks in the detection map, comparing them to the general background spectrum to validate whether a detection is real.  If this were implemented at this point, e would certainly have been flagged and discovered without data from other observations, because its spectrum is markedly different from the background spectra at the same angular radius from the star (see \S\ref{spectrasect}).  Another possible way to increase detection probability would be to use an assumed input spectrum, other than white, for the type of companion being sought.  These improvements are discussed in \citet{s4}.

\begin{figure*}[htb]
\center
\resizebox{.8\hsize}{!}{\includegraphics[angle=0.]{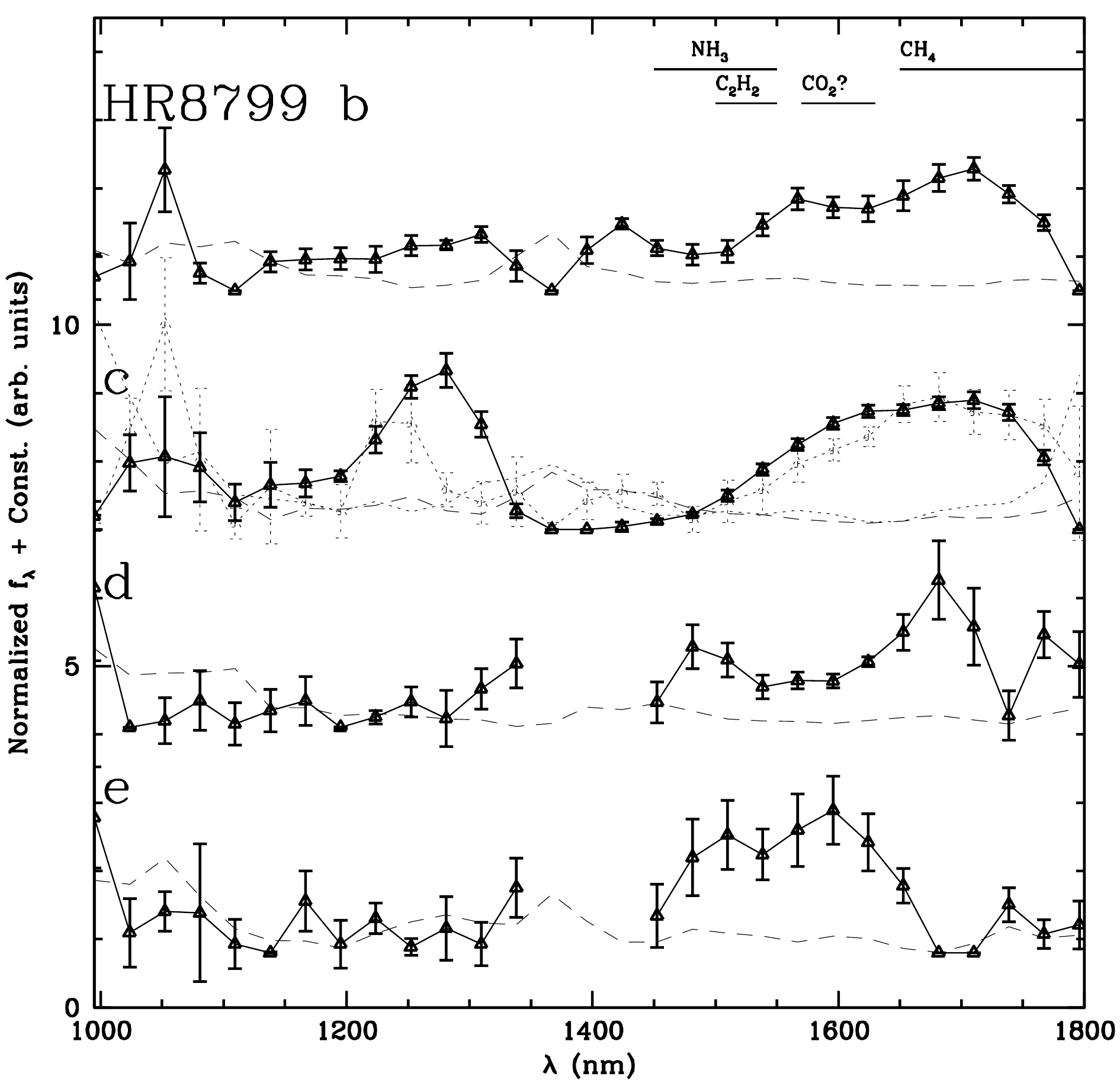}}
  \caption{Spectra of the b, c, d and e components using the S4 algorithm.  1-$\sigma$ error bars are indicated on either side of each point.  Spectra are shown for comparative purposes in normalized f$_\lambda$.  The dotted spectrum of component c is extracted from the October 2012 data and shows poorer detection in the $J$-band due to seeing conditions.  All other spectra are from the June 2012 epoch.  Black dashed lines are spectra of the background extracted at the same radial distance from A for each component.  They consist of the average of 6 different randomly-selected locations in the azimuthal direction; they represent comparative ``noise level'' estimates.  Tentative identification of some molecular features (excluding water) are indicated at the top of the plot.  CO$_2$ is listed with a question mark as explained in \S\ref{molecules} and could also be attributable to HCN. } 
  \label{S4spectra} 
\end{figure*}

\section{Photometry}
From the cubes of speckle suppressed data, photometry can be extracted.  KLIP data have been shown to contain negligible bias in photometry \citep{2012ApJ...755L..28S}, unlike the LOCI algorithm which was used for previous determinations of photometry of this system \citep{mdr00,mdn03,mmb08,mmv10}.  Thus it is the best choice for deriving photometry here.  The data are in units of contrast, so they can easily be converted into photometric measurements using the absolute magnitude of the primary star \citep[M$_{\rm H}$ = 2.30;][]{csv03}. We estimate the broadband photometry by propagating these spectra through the $J$  and $H$ MKO filter bandpasses. For the channels where a detection cannot be established we the value of the spectrum is not included in the photometry. Our findings are summarized in Table \ref{photom} and are in good agreement with published values, albeit with slightly lower $J$-band flux than reported in \citet{mmb08} for HR~8799 c.  Interestingly, this makes c closer in color to b and d.  The d and e components were not detected in the $J$-band.  

We note that all data from each epoch were needed to make these measurements and there is no statistically significant difference between the photometry for c in June and October.

\section{Spectroscopy}\label{spectrasect}

We derived spectra of each of the four point sources using both KLIP and S4, with neither algorithm making any assumptions of spectral shape.  Both algorithms yield consistent results for the brightest detections in the wavelength range.  These are the critical wavelength channels for the photometry presented above, meaning that either algorithm could be used for photometry, but due to the calibration of the KLIP output, it was more straightforward to use KLIP data.  For the fainter channels, the S4 algorithm is superior at obtaining the spectra, because it uses all of the information in the original data cubes---KLIP treats the channels independently and thus does a poorer job of speckle suppression.  For these reasons we decided to conduct spectral analysis on the S4 extractions, which are presented in Fig.~\ref{S4spectra} in arbitrary, normalized units of f$_\lambda$, which is commonly done for comparative spectroscopy.  The spectrum of the c component from the October 2012 observations is also shown in Fig.~\ref{S4spectra}.  It is consistent with the June 2012 spectrum in the $H$-band within the error bars, except for a minor 2-$\sigma$ difference in a new feature in the points flanking 1600 nm (see \S\ref{interp}).  We are unaware of observations that would correlate this difference with variability in the primary star, such as a large increase in UV flux that might induce  photochemistry to produce a new absorption feature.  In the October 2012 data the c component was only marginally detected in the $J$-band due to the poor seeing conditions.  Thus the apparent, but not real, discrepancy between June 2012 and October 2012 at roughly 1300 nm.  All other points are consistent within the errors.   For components d and e, we have excluded the four points obtained within the telluric water band between 1350 and 1430 nm, because what little astronomical signal is present is dominated by starlight.  For b and c this was not an issue because of their much more distant separation from the star.  The other channels had negligible cross-talk from the star, although the points short ward of 1350 nm for e and d are considered upper limits in our analysis.

As indicated in Appendix \ref{s4sect}, the exact range of principal components and size of regions used depends upon seeing and other weather conditions for optimal SNR.  The extractions for October required a larger number of principal components.  (See Appendix \ref{s4sect}.)

\section{Tests of Spectral Extraction Fidelity}\label{tests}

Because these objects have only been studied spectroscopically in a limited manner \citep[see \S\ref{prevspec} ][]{bld10,bmk11} there is only one part of the b component's spectrum to compare with to demonstrate consistency.  In addition, the techniques we used to detect and measure the spectra are new.  Therefore, we conducted several tests to provide additional verification that the spectra are not contaminated by some aspect of the instrumentation or the algorithms employed. We note, however, that Project 1640 has published numerous spectra of much brighter companions (ones that do not require speckle suppression to detect) and these are all consistent with other methods of near-IR spectroscopy.  See, for example \citet{zoh10}, \citet{hmo11}, \citet{hob10}, \citet{rrb12}, \citet{2012ApJ...757...57P} and \citet{2013ApJ...763L...9H}. In addition a Project 1640 disc-integrated spectrum of Titan agrees with other published spectra in this wavelength range, exhibiting strong methane and water absorption.  The fact that bright companion spectra are reliably extracted does not mean that the same techniques can be applied to faint companions below the speckle floor.  Thus we present four separate tests below of the S4 spectral extraction method (Appendix \ref{s4sect}), and some statistical comparisons of the four spectra.

We also note that because both KLIP and S4 reproduce the same features within the errors (Fig.~\ref{klipvss4}) in the spectra and are different techniques, the confidence in the spectral extractions is already high.

\begin{figure}[htb]
\center
\resizebox{1.0\hsize}{!}{\includegraphics[angle=0.]{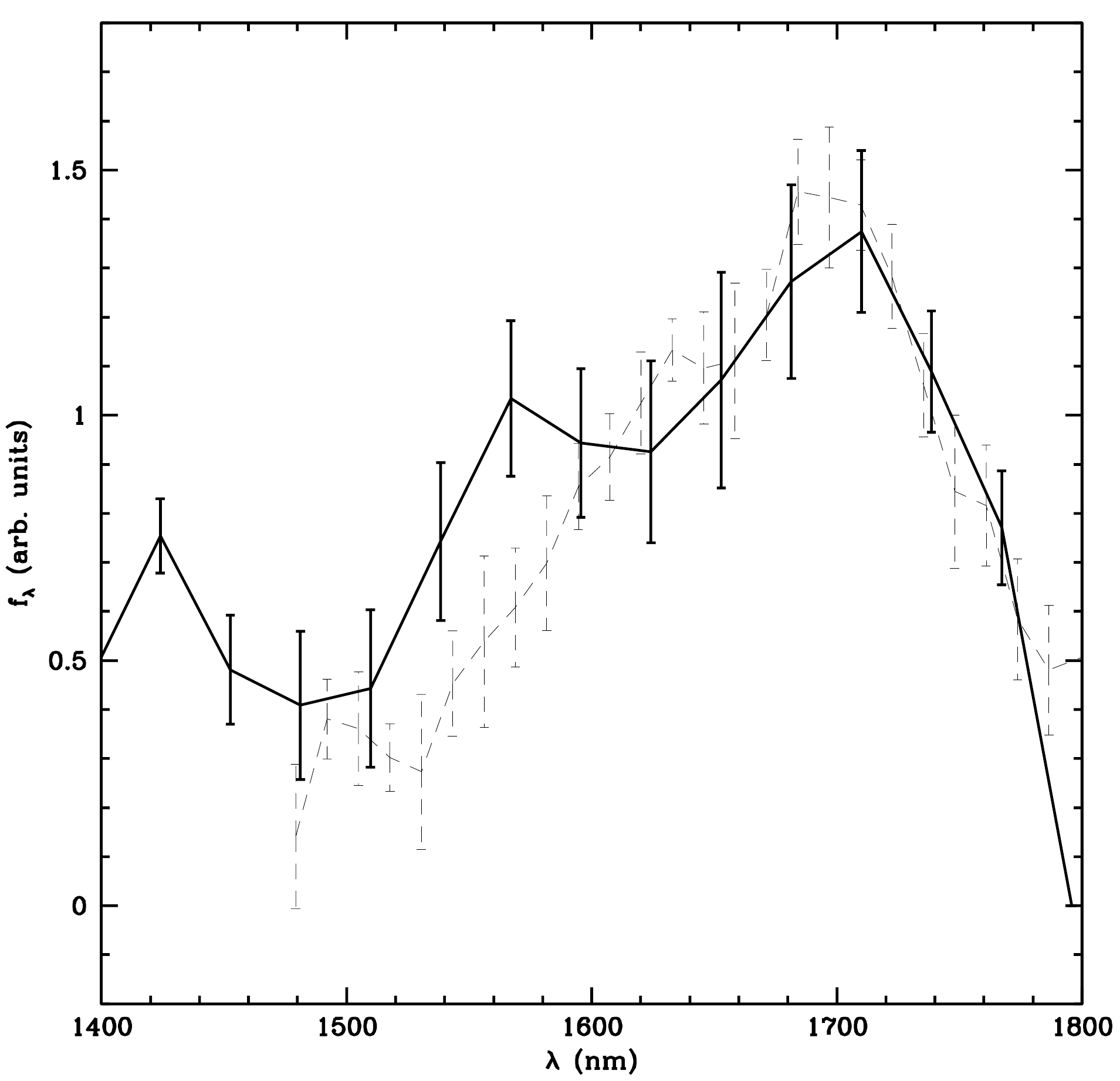}}
  \caption{Spectrum of HR~8799b.  Solid line is our work and the dashed line is from \citet{bmk11}.  The two spectra are consistent within the error bars, except for two points at $\lambda = 1538$ and 1567 nm, which deviate by 2.1 and 2.8-$\sigma$.  The spectra are both shown in normalized units of f$_\lambda$. }
\label{barman} 
\end{figure}

The statistical significance of the detections, in the form of an average over all $\lambda$ for b and c and just $H$-band for d and e, and a maximum, both in units of signal-to-noise ratio, using the background spectrum as the baseline noise against the error bars, is listed in Table~\ref{sigma}.  We note, however, that our methods of determining the error bars and using the background spectrum as the noise level are conservative, intentionally, leading to over estimates of the errors and underestimates of the SNR \citep[see Appendices and][]{s4}.  This statement is bolstered by the data shown in Figs.~\ref{barman}, \ref{compare} and \ref{klipvss4}.

 \begin{deluxetable*}{cccccc}
\tabletypesize{\scriptsize}
\tablecaption{Average and Maximum Signal-to-Noise Ratios\tablenotemark{a}}
\tablewidth{0pt}
\tablehead{ \colhead{} & \colhead{b} & \colhead{c} & \colhead{c} & \colhead{d} & \colhead{e}  \\
 & & June & October & & }
\startdata
Average & 7.0 & 7.9 & 4.2 & 5.1 & 4.3\\
Max. & 15.4 & 29.1 & 9.5 & 9.7 & 8.9
\enddata
 \tablenotetext{a}{SNR values use the background and the error bars. For the last three columns only points in the $H$-band are included.}\\
 \label{sigma}
 \end{deluxetable*}

\subsection{Comparison with Other Studies}
The only companion object to HR~8799 with published spectral information in this study's wavelength range is component b \citep{bmk11}.  Our spectrum in the same wavelength region agrees with the \citet{bmk11} spectrum, except for two points which deviate by roughly 2-$\sigma$.  Fig.~\ref{barman} shows both spectra overlaid to demonstrate the consistency of the two results.  The current spectrum reveals two distinct features however.  These are discussed in detail in \S\ref{molecules}.

\subsection{Other Locations in the Detection Map}
To evaluate the S4 spectral extraction fidelity, for each of the four components, we also extracted spectra at 6 different, randomly selected locations at the same angular radius from the central star.  These spectra are essentially flat and do not reveal the detection of any object.  Rather they indicate the background against which the planets themselves are being detected.  These are shown in Fig.~\ref{S4spectra} by the dashed lines for each component.  Note that in the case of d and e, these background spectra show that the components are only barely detected in the $Y$- and $J$-bands.  However, integration of these channels permits the photometry for b and c and the upper limits for d and e.

Along these lines, we also derived spectra of fourteen other similarly bright peaks in the detection map shown in Fig.~\ref{detmap}, to see whether they were real sources.  All of them, which are either not point sources or of lower significance than the bona-fide companions, either showed peaks in the insensitive water-band between 1350 and 1430 nm---indicating spurious starlight contamination---or flat, featureless spectra at levels similar to the nearby background, indicating that they are not real sources.

\begin{figure*}[htb]
\center
\resizebox{0.85\hsize}{!}{\includegraphics[angle=0.]{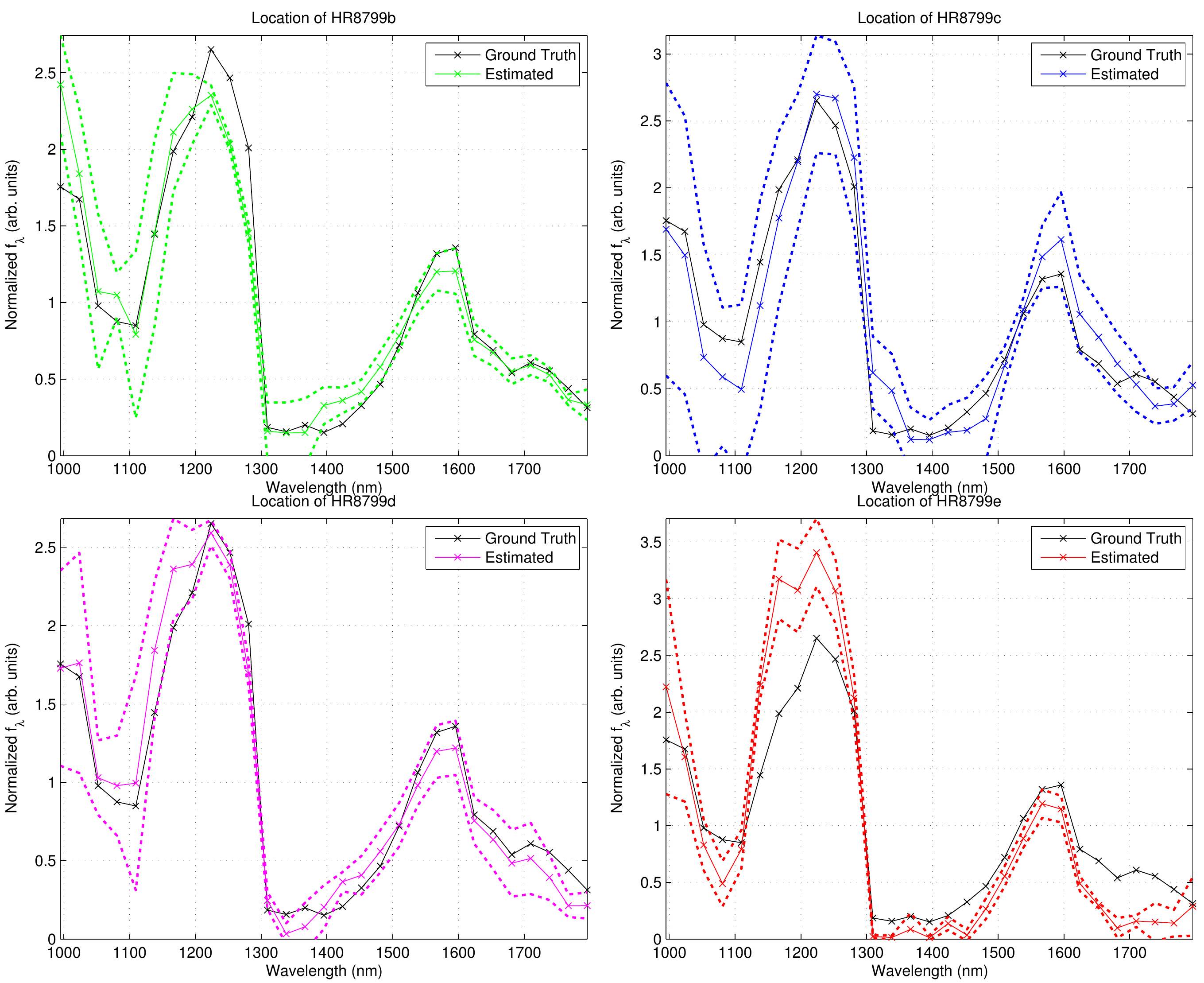}}
  \caption{Spectrum of the T4.5 standard 2MASS J0559-1404 \citep[solid black line, labeled ``ground truth;''][]{2006ApJ...645.1485B} down-resolved to the Project 1640 resolution and used as a test of the spectral extraction technique.  A fake source was injected into the data with this spectrum at the same radial location and intensity as each of the planets in the HR~8799 system.  The retrieved spectrum after the S4 extraction is shown as the colored solid line and with the colored dotted lines indicating the 1-$\sigma$ error bars.  Each extraction is consistent with the original spectrum in slope and individual features (see \S\ref{tests}).  This demonstrates the high fidelity of the S4 spectral extraction technique.}
\label{fakeplot} 
\end{figure*}

\subsection{Fake Source Spectrum Retrieval}\label{fake}
As another test of the ability of our data analysis to retrieve spectra with high fidelity we inserted multiple fake companion sources into the same data cubes we analyzed for this paper.  The sources were placed at the same overall brightness as each component and at the same radii from the star.  The fake sources were given the T4.5 spectrum of the standard 2MASS J0559-1404, chosen for its strong molecular features \citep{2006ApJ...645.1485B}, which are useful to understand whether the features are correctly reproduced.  The spectrum was resampled at Project 1640's resolution and wavelength range and applied to a cube of a fiducial PSF derived from an unocculted observation of the internal calibration white-light source.  The instrumental response of the system was also applied to this PSF and then it was reduced in intensity by a factor sufficient to make it of the same intensity as each of the b, c, d and e components.  We then ran this composite data cube through the S4 algorithm, detected each object with the same significance as the real components and extracted their spectra as described in Appendix \ref{s4sect}.  This was repeated 5 times with different, randomly selected azimuthal locations for the fake objects at the same radii as the planets.  

The spectra of fake sources extracted matched the input spectrum, with an rms error of less than 1-$\sigma$ for all four cases.  All four spectra of the fake sources are shown in Fig.~\ref{fakeplot}.  In general the extractions are identical in shape to the input spectrum and in all cases reproduce small kinks in the input T4.5 object.  For reference, each fake component was set to the same 2 to 3\% of the local speckle brightness as the real planets are (\S\ref{instrument}; i.e.~the companions are roughy 40 times fainter than the local speckle background).  The average deviation from the input spectrum, over all wavelengths, is 1.6\%, 8.8\%, 5.6\% and 15.0\% for b through e respectively, consistent with the photometry error values given in Table \ref{photom}.  The largest discrepancies appear in the fake source placed at the radial distance of e, as expected, due to its close proximity to the star.  In particular the $J$-band peak is over-estimated, although the broad-band shape and slope of the spectra are retrieved to better than 20\% accuracy.  We note that all molecular features are reproduced, except for the very reddest part of the $H$-band, which is noisy in any case (cf.~Fig.~\ref{S4spectra}).  The critical point with respect to interpreting the real spectra of the companions is that every minor feature (including, for example, the small dent in the $J$-band peak at about 1250 nm) is reproduced with fidelity in the simulated source spectra.  Generally $Y$ and $J$-band data are worse, because of poorer AO performance at those wavelengths.

 \begin{deluxetable*}{ccccccc}
\tabletypesize{\scriptsize}
\tablecaption{Statistical Covariance for All Pairs of Spectra}
\tablewidth{0pt}
\tablehead{ \colhead{b and c} & \colhead{b and d} & \colhead{b and e} & \colhead{c and d} & \colhead{c and e} & \colhead{d and e} & c and c\tablenotemark{a}  }
\startdata
0.152 & 0.015 & -0.050 & 0.045 & 0.027 & 0.003 & 0.849
 \enddata
 \tablenotetext{a}{For all but the first column only points in the $H$-band are included. The last column is the comparison of c between the two epochs of observation (June 2012 and October 2012).}\\
 \label{correla}
 \end{deluxetable*}

\subsection{Low-Order Bias Tests}
Because of the possibility of apparent systematic, but broad-band, over, or under, estimates of the peaks especially for the radial location of e, as described in the last section, we investigated the behavior of injected fake sources with perfectly white spectra positioned at various locations in the field of view.  This test was designed to reveal any field-dependent biases in the spectral extraction by S4 and to ascertain whether a further minor correction to the spectra extracted is necessary.  No such correction was deemed necessary, as we explain below.

\begin{figure}[htb]
\center
\resizebox{0.85\hsize}{!}{\includegraphics[angle=0.]{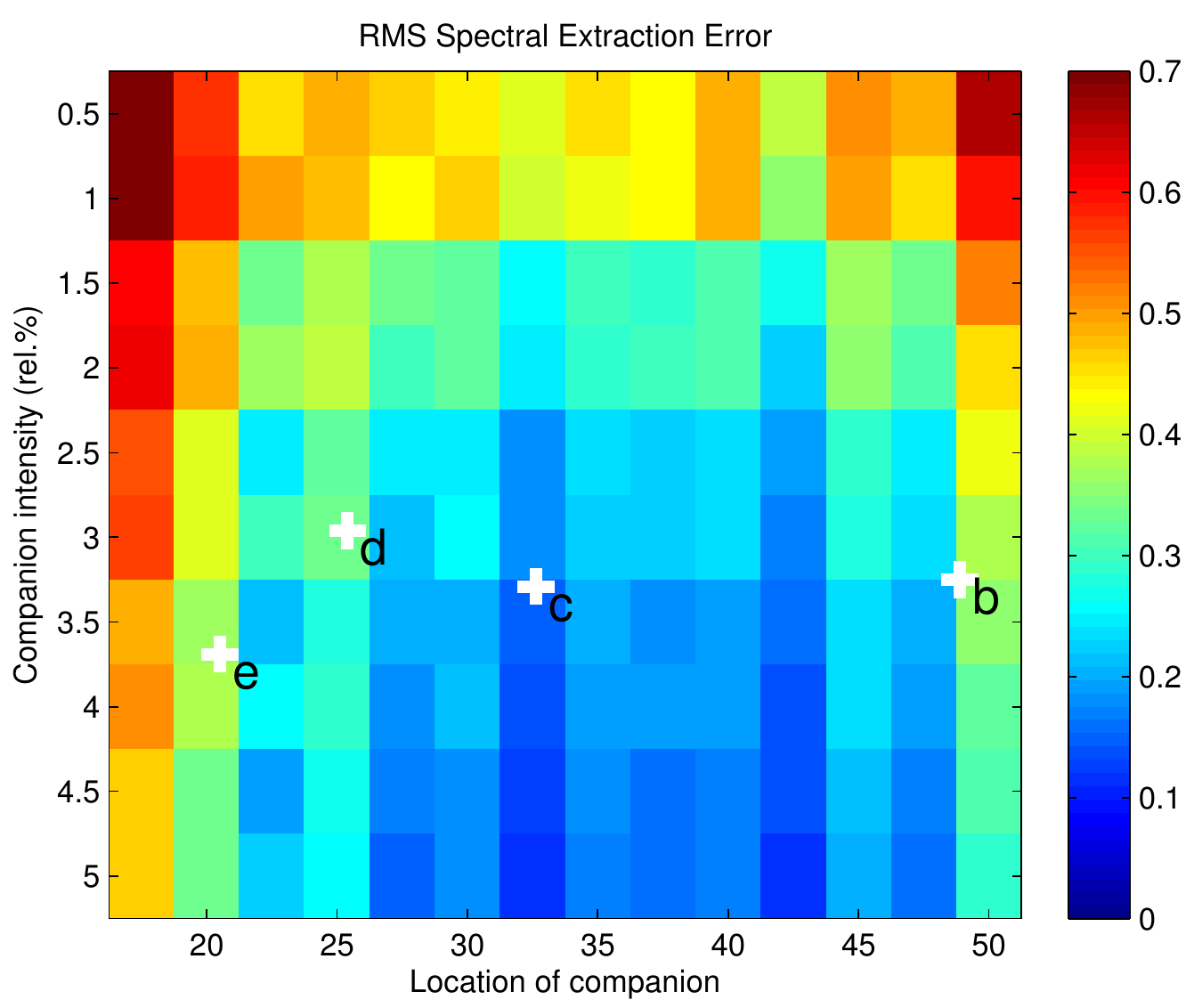}}
  \caption{Surface plot showing the rms error on extracted white spectrum sources injected into the data at a range of intensities (0.5\% to 5\%) relative to the local speckle brightness in the unprocessed data, and a range of radial distances from the star (in pixels, ranging from 15 to 50).  The positions of the detected planets in this parameter space are shown by white crosses and labeled. This test demonstrates the high fidelity of the S4 spectral extraction technique.}
\label{specerrplot} 
\end{figure}

We injected fake white-spectrum companions at a range of intensities (expressed as a fraction of the local speckle intensity) and at a range of radial distances from the star (from 15 to 50 pixels, in steps of 5 pixels). For each radial location we injected fake sources with a range of  0.5\% to 5\% intensity relative to the local speckle brightness, with steps of 0.5\%.   For each of these 140 fake sources, we retrieved the spectrum, as we did with the tests in \S\ref{fake}, and computed the rms error between the injected and retrieved spectra as well as the difference in slope between the two.  None of the extracted spectra had a linear slope that deviates from white by more than 0.2 rms.  To visualize the results of this test, we present a surface plot of the rms error versus radial location and brightness Fig.~\ref{specerrplot}.  The rms error at the locations of each of the four planets is less than 0.4 in all cases, consistent with the tests utilizing a T4.5 spectrum (\S\ref{fake}).  No location in the explored parameter space exceeds an rms error of 0.75.  

Since no systematic error was detected, and no additional correction of spectral shape is required, we conclude that the spectra shown in Fig.~\ref{S4spectra} are real and as accurate as the data permit.  Thus, it is appropriate to interpret them.

\section{Observational Interpretation of the Spectra}\label{interp}
 
 Even in comparison with other very low temperature objects, such as T-dwarfs, the spectra in Fig.~\ref{S4spectra} are, first, all different from each other and have significant differences from known objects.  The HR~8799 companions all exhibit red colors in the near IR, while the lowest temperature brown dwarfs maintain relatively blue colors in $J$  and $H$-bands.  However, the newly discovered examples of so-called ``Y'' dwarfs, though likely much cooler than the objects studied in this paper, may be red as shown by \citet{2012ApJ...759...60T}, \citet{2012ApJ...758...57L}, \citet{2012ApJ...756..172M} and especially \citet{2012ApJ...753..156K}.  In fact it has been suggested in some of these papers that as sensitivity to very low temperature or low surface gravity objects increases, instruments may need to be built to observe at increasingly longer wavelengths, with the $J$  and $H$-bands becoming so under-luminous that the objects of interest fall below detection limits.  Indeed, the d component exhibits a $J$$-$$H$ color of $>2^{\rm m}$, whereas a mid-T dwarf has a $J$$-$$H$ color of 0.0. However, it is important to note that some of the warmer L-dwarfs can have colors that are this red in the near-IR.
 
 \subsection{Are the Four Spectra Statistically Different?}
As a very rudimentary initial measurement, we calculated the statistical covariance among the six possible pairs of four spectra and the two spectra of c, to establish how similar or dissimilar they are from each other.  Except for the comparison of the two epochs of the observation of c, no two have a covariance above 0.8 (a common measure of a real correlation) and all are below 0.2, a general benchmark for a significant statistical difference.  Components b and c have a covariance of 0.152, which also indicates significant statistical difference between the two objects, though their covariance is higher than in the other cases.  The calculated values are tabulated in Table~\ref{correla}.  For the calculations in which d or e were involved, only $H$-band data were used, since the $Y$ and $J$ spectra are only marginal detections.

The covariance values suggest that a very weak similarity exists between b and c (essentially the main hump in the $H$-band). These values also suggest that the others are all significantly different from each other, confirming an inspection by eye of Fig.~\ref{S4spectra}.
 
\begin{figure*}[htb]
\center
\resizebox{1.0\hsize}{!}{\includegraphics[angle=0.]{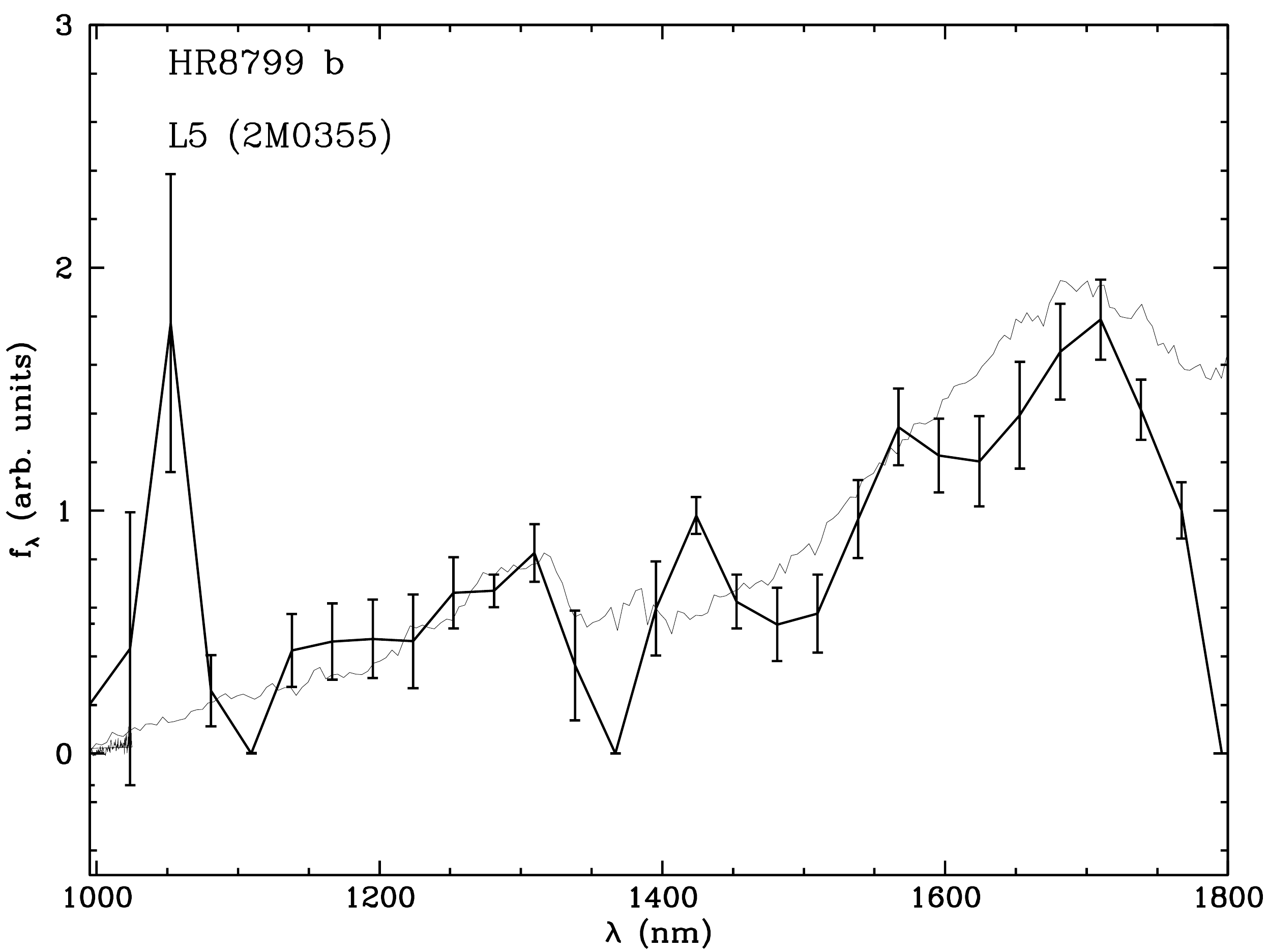} \includegraphics[angle=0.]{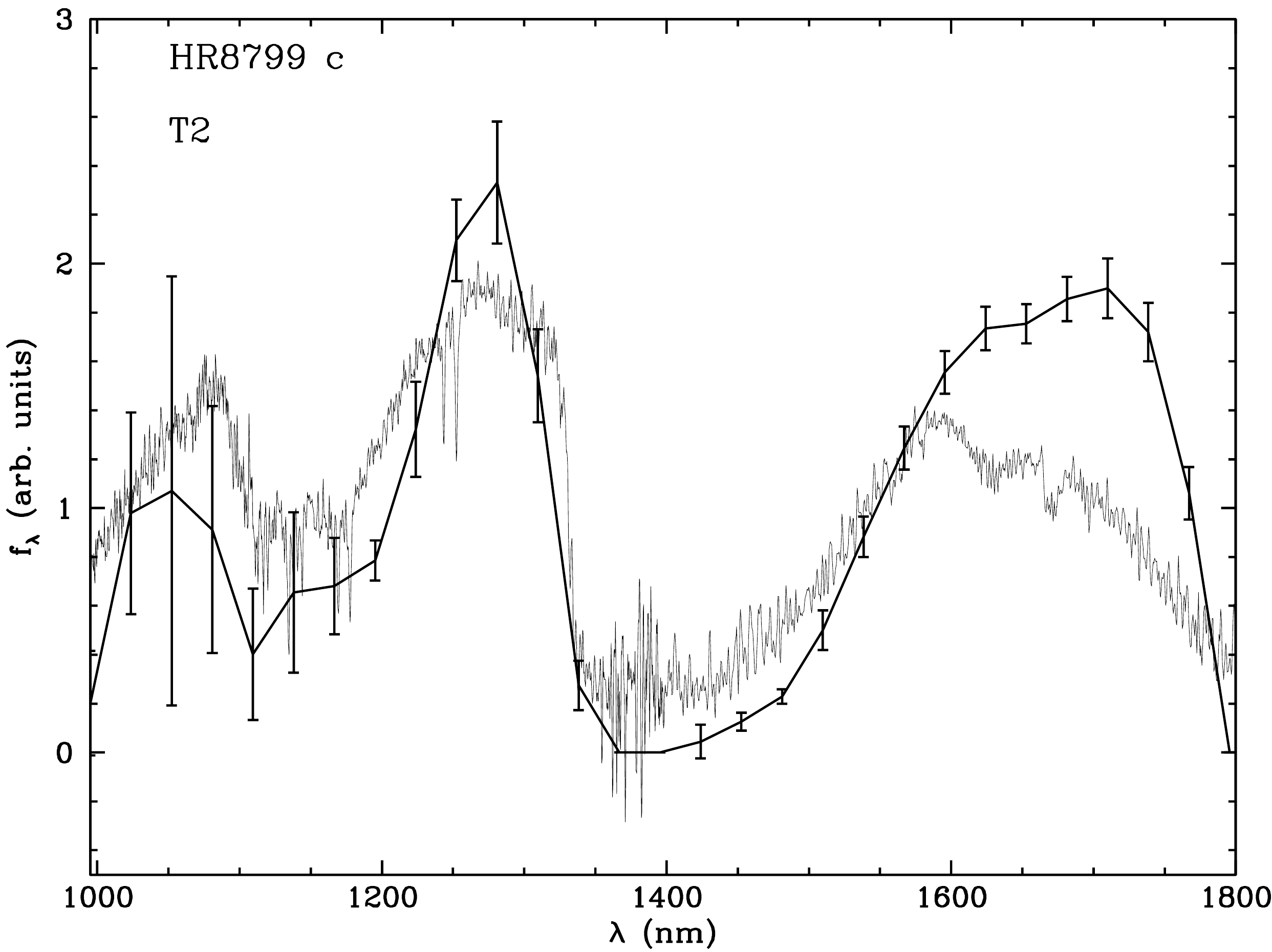}} \newline
\resizebox{1.0\hsize}{!}{\includegraphics[angle=0.]{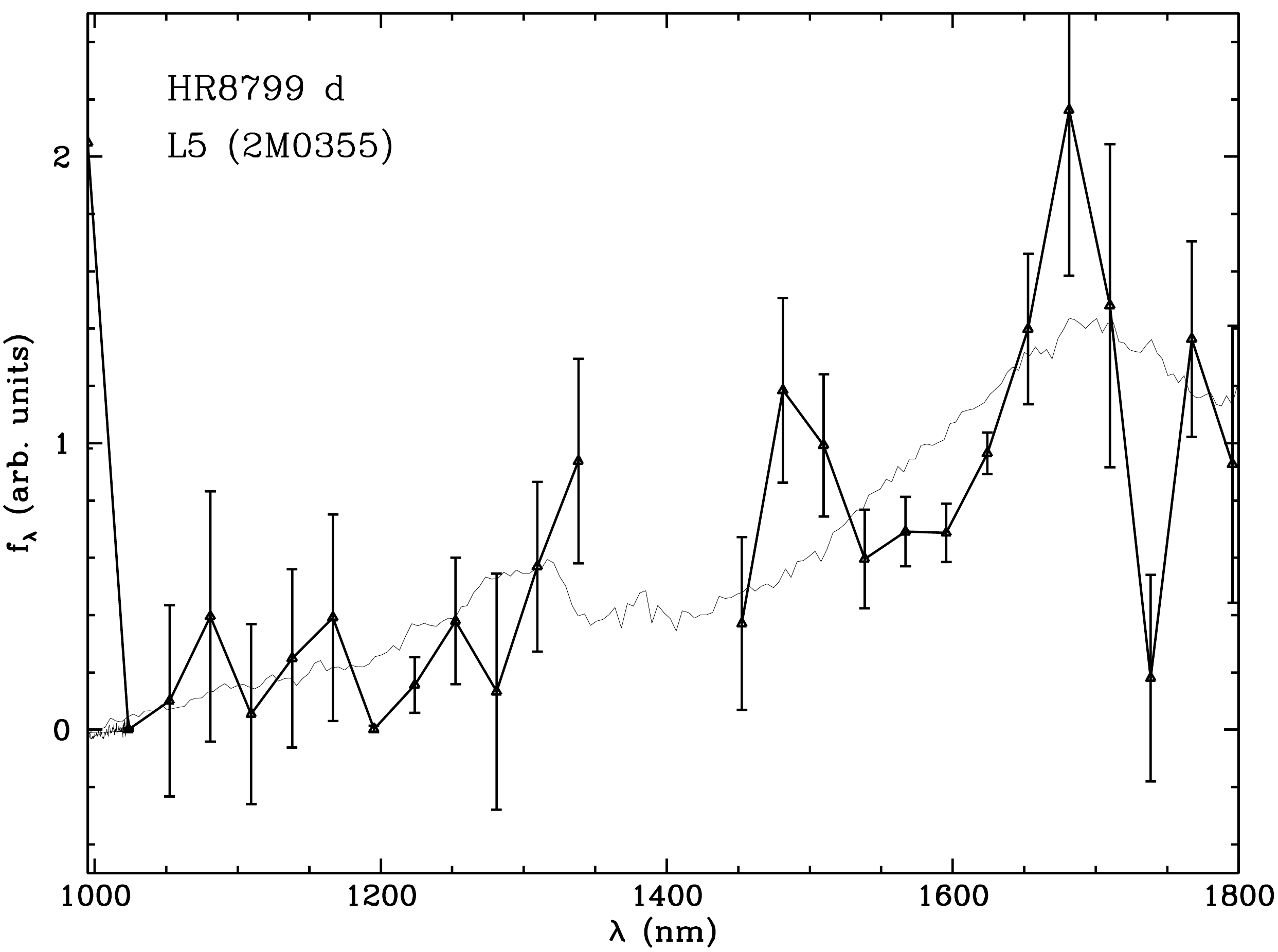} \includegraphics[angle=0.]{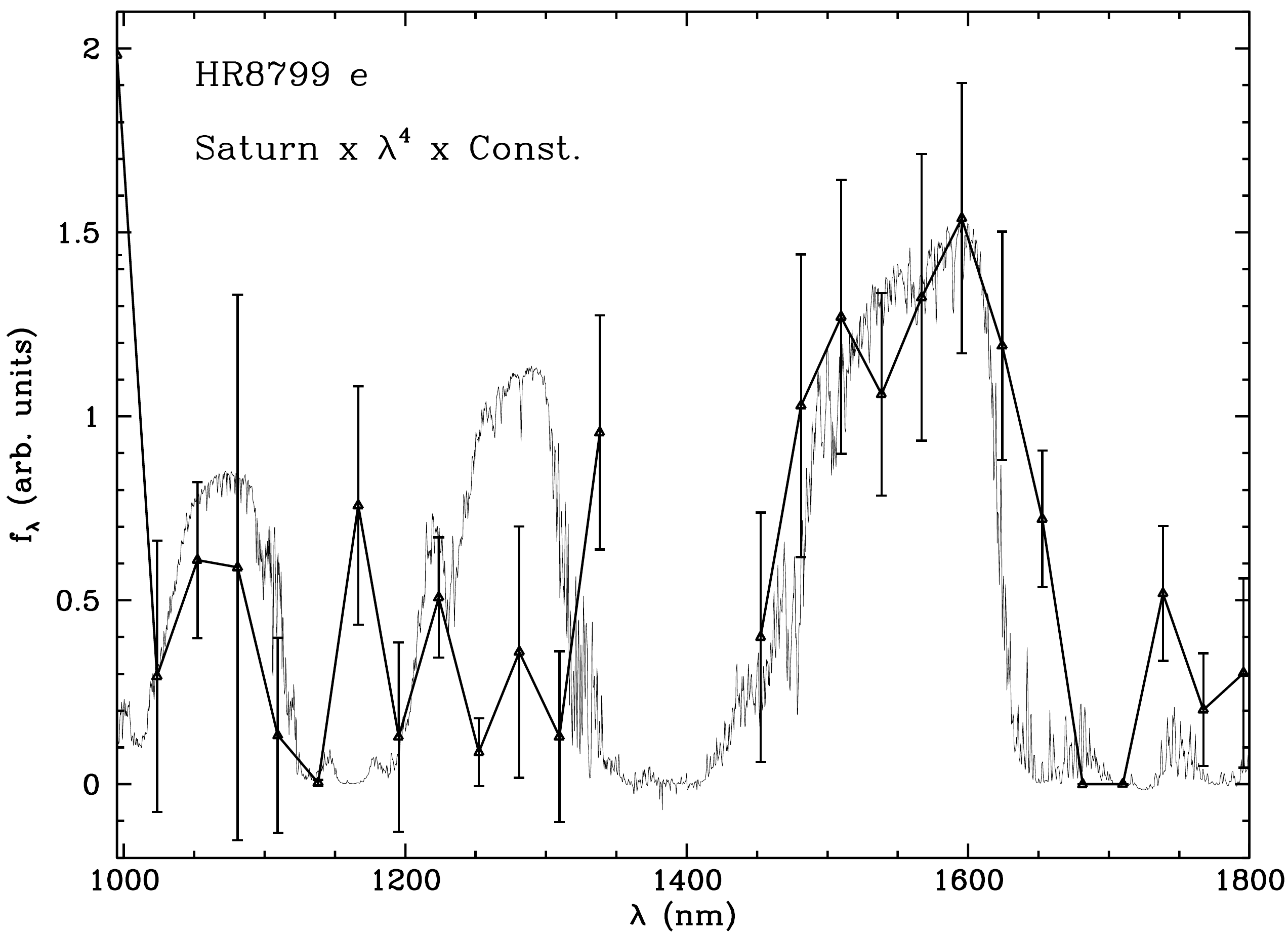}}
\caption{Spectra of HR~8799 b (top left), c (top right), d (bottom left) and e (bottom right) with comparison objects.}
\label{compare} 
\end{figure*}

\subsection{Comparison with Other Objects}
For comparison purposes, we attempted to find spectra of other celestial point sources that exhibit at least some similarity to those of the components of HR~8799.  We primarily used the library of spectra in \citet{rcv09}.  Initially we plotted the four spectra overlaid with those of the outer planets of the solar system.  However, the planets do not provide reasonable matches, with the sole exception of the e component, as described below.  We expanded the pool of comparison objects to stars and brown dwarfs.  In Fig.~\ref{compare} we show comparisons with objects of best match presented for each of the four companions of HR~8799.  In all cases there are discrepancies, which clearly shows that these four objects are currently unique.  Sources b and c seem to exhibit the usual water features between $Y$, $J$  and $H$-bands, while for d and e we only detect the red edge of the water band between $J$ and $H$-band.  The main methane feature at 1.65 $\mu$m seems present only in the d and e components, and strongest in e.  See \S\ref{molecules}.


In Fig.~\ref{compare} (top-left), we overlay the spectrum of the peculiar L5 brown dwarf 2MASSJ035523.51+113337.4 \citep[2M0355;][]{2013AJ....145....2F} with the b component of HR~8799.  This brown dwarf, sometimes referred to with a L5$\gamma$ spectral type, seems best explained with low gravity and a possible reddening due to dust \citep{2013AJ....145....2F}.  Although the general shape of the spectrum of 2M0355 may be a match, none of the small features are reproduced.  We believe these to be molecular features and they are discussed in \S\ref{molecules}.  

In the top-right panel of Fig.~\ref{compare}, we compare HR~8799c  to the T2 dwarf SDSS J125453.90-012247.4.  Clearly this object is not completely similar to a field T dwarf, as significant discrepancies appear at many wavelength channels.  The comparison T dwarf is somewhat bluer with HR~8799c showing a lower brightness in the $Y$-band, and the lack of methane in the $H$-band, while the $J$-band peak appears sharper, or ``pointier.''  We note that the covariance of these two spectra is $-0.41$ which is dominated by the broad slope difference.

In the bottom-left panel of Fig.~\ref{compare} we overlay the spectrum of 2M0355 \citep{2013AJ....145....2F} with the d component of HR~8799, as we did in the top-left panel for the b component.  Important discrepancies exist in several features around 1500 nm and 1660 nm in the d spectrum.  These are discussed further in \S\ref{molecules}.

The bottom-right panel of Fig.~\ref{compare} shows perhaps the best match between another object and one of the HR~8799 planets, the e component.  Overlaid is the spectrum of Saturn, but reddened by a function of $\lambda^{4}$ and normalized to match the overall flux level.  This reddening function is really an attempt to approximate the spectrum of Saturn with the Sun's Rayleigh-Jeans black body spectral tail removed, as though it were a night-side spectrum of Saturn.  We are unaware of any night-side spectra of the giant planets in this wavelength range, and so employed this simplistic approach.  We note that there is a strong indication of methane in this object.  

Recently, \citet{2012ApJ...754..135M} and \citet{2012ApJ...756..172M} have produced new models that incorporate far more complex and realistic cloud and condensate behavior.  These have been applied to the previously published photometry of the HR~8799 system and indicate that clouds at varying levels in the atmosphere and with incomplete covering fraction of the atmosphere (as is seen in the giant planets of our solar system) may explain the red behavior of the spectral energy distributions.  The new data presented here should provide additional constraints to these models.  

\subsection{Initial Identification of Molecular Features}\label{molecules}
Using the database of \citet{2008ApJS..174..504F} and in personal communications with M. Marley and D. Saumon,
 we have indicated in Fig.~\ref{S4spectra} the four main sources of opacity that most closely match the features we see in these four objects (aside from the water opacities between the astronomical bandpasses).  These include ammonia, NH$_3$, in the 1450 to 1550 nm range, acetylene, C$_2$H$_2$, in the 1500 to 1550 nm range, methane, CH$_4$, redward of 1650 nm, and possibly CO$_2$ in the 1560 to 1630 nm range, a feature which is commonly used in terrestrial atmospheric science \citep[e.g.][]{OBrien:2002fk}.  Given the temperatures and masses of these objects as discussed in \S\ref{intro}, these seemed the simplest molecules to consider. In Fig.~\ref{S4spectra} we indicate the CO$_2$ feature with a question mark because it is unclear that it should be visible at the high temperatures inferred for these objects.  

Based on the spectra, and assuming we have identified the sources of opacity correctly, we propose the following: 
\begin{itemize}
\item b: contains ammonia and/or acetylene as well as CO$_2$ but little methane.  
\item c: contains ammonia, perhaps some acetylene but neither CO$_2$ nor substantial methane.
\item d: contains acetylene, methane and CO$_2$ but ammonia is not definitively detected.
\item e: contains methane and acetylene but no ammonia or CO$_2$.
\end{itemize}

We emphasize here that these are tentative identifications, although the lack of methane in the b component is consistent with the work in \citet{she12}, \citet{bmk11} and \citet{bld10}. In contrast, \citet{bmk11} identify CO, not CO$_2$ as we indicate here.  Detailed modeling by theorists will be necessary to definitively test these identifications.  

Acetylene, has never been convincingly identified in a substellar object outside the solar system, and unfortunately the fundamental band is in the thermal infrared, for which we have no spectral coverage.  Therefore it is difficult to confirm with other observations at this point. 

\subsection{Variability and Source of Some Molecules}
As mentioned in \S\ref{spectrasect}, there is a weak 2-$\sigma$ difference in the c spectrum between June 2012 and October 2012 that is suggestive of the presence of a weak CO$_2$ feature at $\sim1610$ nm in October 2012 that is not apparent in June 2012.  Given the new models with patchy clouds, which imply variability in these objects, it is not a leap of faith to consider variability of certain molecular features in these objects.  It is probably unlikely that broad-band variability in the primary star (as discussed in \S\ref{intro}) is the cause of such a feature, mainly because the emergent flux from each of these planets is on the order of a few times $10^7$ ergs cm$^{-2}$ s$^{-1}$ (assuming the $\sim 800$K temperatures others have derived) while the incident starlight ranges from 3 to 4 orders of magnitude smaller.  

However, the UV flux from the primary star is considerable because it is an A5V star---with more than 1000 times the UV flux of the Sun.  If the Lyman-$\alpha$ line is also prominent in HR~8799, there may be larger UV radiation incident upon even the most distant b component compared to that upon Jupiter.  As \citet{2009arXiv0911.0728Z} demonstrate, UV flux combined with strong mixing in the atmosphere can greatly complicate the chemistry represented in the emergent spectra of planets (such as Jupiter, in their paper).  Acetylene and even hydrogen cyanide (HCN) can become abundant.  We mention HCN because it has an absorption feature at 1580 nm, comparable to the feature we have tentatively assigned to CO$_2$.  A known FeH feature exists there as well, but has broader extent.

We will continue to monitor the spectra of these planets into the future to assess the reality of any variability.  We note that for years photometric variability of L-dwarfs has been known, and detected in some T-dwarfs as well \citep[e.g.][and references therein]{2013AJ....145...71K}.  From this point of view, it would not be surprising to detect variability in any or all of these planets.

As discussed in \S\ref{intro} and shown in Table \ref{photom}, c, d, and e differ by less than a magnitude in $H$-band, but their spectra are not the same.  Furthermore, b has some similarity in spectral shape when compared with c, which is nearly a magnitude brighter.  This includes the peak near 1350 nm and the general shape of the large feature in $H$-band.  In addition, the e component has similarity with a reddened spectrum of Saturn.  All these facts point to a diversity in salient properties of objects for which mass, age and metallicity are not the only factors determining observables.  

Our results also highlight the power of discerning spectral differences between planets in multi-planet systems, which can provide constraints on the formation locations for each object. Specifically, \cite{2011ApJ...743L..16O} point out that the measured carbon-to-oxygen ratios in exoplanet atmospheres may constrain the formation locations of such objects, since various snow lines of carbon and oxygen-rich ices form at various radial locations from the star.  

The results presented here consist of the first comparative spectroscopic study of multiple planets around a star other than our Sun.

Finally we note that it would be of tremendous value to the exoplanet community for any space missions passing behind any of the planets of the solar system to take night-side thermal spectra of the planets in the near IR, to enable direct comparisons with exoplanets.

\section{Discussion: Planets are Diverse}
 We have attempted to provide significant new material in the understanding of the exosolar system of HR~8799, through spectroscopy of all of the known planets.  What is most striking is that previous studies of the system indicate that components c, d and e are roughly the same near-IR luminosity (within roughly a magnitude).  However, as we have shown, their spectra are substantially different from each other.  What causes this remains unknown.  Whether this is due to small differences in formation, metallicity differences as a function of orbital radius, or evolutionary differences all remain questions.  On the other hand, it is important to note that these are the first spectroscopic observations of multiple planets in a planetary system other than our own.  Thus it is, perhaps, not surprising that four planets that differ by only a magnitude or two in brightness exhibit such diversity in their spectra.  Furthermore, similar rapid spectral changes happen at the L/T and M/L transitions. 
 
 The spectra of HR~8799 b, c, d and e indicate much redder colors than objects with similar spectral features (such as methane) currently known.  Some authors have provided explanations for the extremely red nature of the spectra through far higher cloud content than previously thought for objects in this mass and temperature range.  In addition, a cloud covering fraction smaller than 100\% may be exhibited in these objects \citep{she12,cbi11,2012ApJ...754..135M,2012ApJ...756..172M}.   
 
Given the portrait of the star and other aspects of the system described in \S\ref{intro}, we note, in particular, the extreme variability of the star in the context of our own solar system.  The primary star can vary by as much as 8\% in $V$-band luminosity over a period of two days \citep{1999MNRAS.303..275Z}.  Compared to the Sun, this is a violent environment for the stability of any planet's atmosphere.  The solar constant exhibits variability on the order of several parts per million \citep[e.g.][]{1987JGR....92..796F,2012SoSyR..46..170F}, and the variability due to the Earth's non-circular orbit is on the scale of 6\%, but over 6 months, not two days \citep{2012SoSyR..46..170F}.  In the case of HR~8799 variability of the star is 3 orders of magnitude greater over a hundred times shorter timescale.  Furthermore, given the A5V spectral class, the planets in this system must be irradiated with far larger fluxes in the aerosol-generating UV wavelength range.  If complex UV photochemsitry is present, the planets may exhibit variability in spectral features.  While our initial detection of such variability in CO$_2$ or HCN in the c component is of such low confidence that it cannot be claimed definitively, future monitoring of the spectra of these objects is warranted.

\section{End Note: Project Context and Survey}

This communication represents the culmination of over a decade of work by the authors, including efforts in science-grounded instrument conception and design; optical, mechanical and electrical engineering; development of novel techniques for the manipulation and control of light from distant stars at the level of $\lambda/1000$; systems engineering and integration; control and data reduction software, which, in this case, comprises several large efforts that should be considered ``instruments'' in their own right; software for the identification and spectrum extraction of possible companions---an effort that includes expertise from the field of computer vision; advanced detector control; and all of the tools of modern astronomy brought to bear on the fundamentally difficult problem of high-contrast imaging: astrometry, coronagraphy, spectroscopy, photometry and various aspects of point source analysis and signal processing.

This paper marks the beginning of a three-year survey to search for and characterize new exoplanetary systems around the nearest A and F stars.  With the advent of this new project and several others beginning operations in the next two years, comparative exoplanetary science is beyond the initial technical hurdles and can now move into a stage of exploring the range of planets extant.  The techniques will, of course, continue to improve, and an exciting research field lies ahead.

\acknowledgments
 We are especially grateful for our two referees both of whom were particularly meticulous, thorough and fair.  Their reports resulted in significant improvements in the paper.  A portion of this work is or was supported by the National Science Foundation under Award Nos.~AST-0215793, 0334916, 0520822, 0619922, 0804417, 0908497, 1039790 and EAGER grant 1245018.  A portion of the research in this paper was carried out at the Jet Propulsion Laboratory, California Institute of Technology, under a contract with the National Aeronautics and Space Administration and was funded by internal Research and Technology Development funds.   A portion of this work was supported by NASA Origins of the Solar System Grant No.~NMO7100830/102190, and NASA APRA grant No.~08-APRA08-0117.  BRO acknowledges continued support from Paco.  Our team is also grateful to the Plymouth Hill Foundation, and an anonymous donor, as well as the efforts of Mike Werner, Paul Goldsmith, Jacob van Zyl, and Stephanie Hunt.  ELR acknowledges support from NASA through the American Astronomical Society's Small Research Grant Program.  RN performed this work with funding through a grant from Helge Ax:son Johnson's foundation. LP performed this work in part under contract with the California Institute of Technology funded by NASA through the Sagan Fellowship Program.  SH is supported by a National Science Foundation Astronomy and Astrophysics Postdoctoral Fellowship under Award No.~AST-1203023.  Any opinions, findings, and conclusions or recommendations expressed in this material are those of the authors and do not necessarily reflect the views of the National Science Foundation.  RF and DWH were partially supported by the National Science Foundation under Award No.~IIS-1124794.  BRO thanks Didier Saumon and Mark Marley, as well as Bruce Macintosh, Gilles Chabrier and Isabelle Baraffe, as well as Jackie Faherty and Statia Cook, for comments and discussion of a draft prior to submission.  We thank the the Raymond and Beverly Sacker Foundation whose generous donation allowed the purchase of the original Project 1640 detector. We also thank Teledyne Imaging Sensors for their help and support throughout this project.  We thank the dedication and assistance of Andy Boden, Shrinivas Kulkarni and Anna Marie Hetman at Caltech Optical Observatories.  Finally, the entire team expresses sincere gratitude and appreciation for the hard work of the Palomar mountain crew, especially by Bruce Baker, Mike Doyle, Carolyn Heffner, John Henning, Greg van Idsinga, Steve Kunsman, Dan McKenna, Jean Mueller, Kajsa Peffer, Kevin Rykowski, and Pam Thompson.  This project would be impossible without the flexibility, responsiveness and dedication of such an effective and motivated staff.


\appendix

\section{The KLIP Speckle Suppression Algorithm}\label{klipsect}

\subsection{Detection}

The extracted data-cubes are first cleaned for remaining spurious bad pixels using median filtering and high-pass filtered to remove the contribution of the residual atmospheric halo. Their radial scaling  and relative registration are then calculated using the procedure discussed in \citet{cpb11} and \citet{pcv12}. For each wavelength from $\lambda_0 = 995$ nm, all of the subsequent slices in each cube of the observing sequence are then compressed or stretched and registered. In these pre-processed cubes all speckle patterns appear at the same spatial scale and the PSF at $\lambda_0$ is the true image of the sky. This yields a series of $N_{\lambda}$ pre-processed cubes of dimensions $N_{\lambda} \times N_{\rm Exposures} \times N_{\rm Spaxels} \times N_{\rm Spaxels}$. We then proceed to speckle removal using the KLIP algorithm described in \citet{2012ApJ...755L..28S}. The target images consist of the $ N_{\rm Exposures}$ PSFs at $\lambda_0$ in the pre-processed cube $n_{\lambda_0}$. We then proceed to partition the images into search zones, $\mathcal{S}$, that are equivalent to the optimization zones in LOCI as presented by \citet{lmd07} (radial location $r$, radial width $\Delta r$, azimuthal location $\phi$, azimuthal width $\Delta \phi$). Note that in contrast to LOCI, for KLIP these are not subtraction zones. For each target image the reference ensemble is chosen as the subset of rescaled slices in the same exposure whose wavelength $\lambda$ is such that the image of a putative planet in the compressed/stretched images is sufficiently far from its location at $\lambda_0$ (parameter  $N_{\delta}$ in \citet{lmd07,cpb11,pcv12}). Finally the $N_{\rm Exposures}$ speckle-reduced images at $\lambda_0$  are co-added and we proceed to the next wavelength. 

We  conduct a Monte-Carlo search over the ensemble of parameters $(N_{\delta}, \Delta r, \Delta \phi,K_{\rm KLIP})$ with $N_{\delta} =0.5,0.75,1,1.25,1.5 \; W$ (W is the un-occulted PSFs FWHM), $\Delta r = 10,20,30$ pixels, $\frac{\Delta \phi}{2 \pi} = \frac{1}{2},\frac{1}{4},\frac{1}{6},\frac{1}{8},\frac{1}{12},\frac{1}{16},\frac{1}{20},\frac{1}{24}$, and $K_{\rm KLIP} = 1.25$. This approach is reminiscent of the parameter search in Soummer et al. (2011). However, when compared to LOCI, it is greatly facilitated by the absence of subtraction in KLIP  and by our moderate number of reference PSFs. HR~8799 b and c are visible in $H$-band reduced images (SNR $>8$) and detected using matched filtering with un-occulted PSFs, at SNR $\sim 3$, in $J$-band reduced images for $\frac{\Delta \phi}{2 \pi} <  \frac{1}{6}$ and $K_{\rm KLIP}> 10$. Since HR~8799 d and e lie at closer angular separations they are detected in $H$-band by matched filtering, at SNR $\sim 3$, for $\frac{\Delta \phi}{2 \pi} > \frac{1}{8}$ (values of $N_{A}$, see \citet{lmd07}, similar to the ones at the location of HR~8799 b, c) and $K_{\rm KLIP} >10$. These findings are summarized in the integrated $H$-band image shown in Fig.~\ref{specklesuppimages}.

\subsection{Spectrum Extraction}

Large biases of the spectral information can arise when estimating the spectro-photometry of faint companions unravelled by an aggressive PSF subtraction routine. In \citet{pcv12} we identified two sources of potential biases in IFS data: self-subtraction from over fitting the companion's signal, and wavelength-to-wavelength cross talk. When using KLIP the former can be calibrated using the forward modeling methodology discussed in \citet{2012ApJ...755L..28S}, provided that the self-subtraction is not too severe (e.g. provided that it does not radically  change the morphology of the PSF). However since our detection pipeline relies on search zones of large radial extent, some companion flux is present in the basis-set of principal components. Choosing a small $K_{\rm KLIP}$ can, in principle, mitigate the wavelength-to-wavelength cross talk, because the contribution of the companion's flux to the reference PSF is most likely in the small eigenvalue components.  However, this does not fully alleviate this phenomenon. The full details of this characterization pipeline will be reported in a subsequent paper focusing on the astrometric characterization of the HR~8799 using Project 1640. Below we only outline the main steps of our method.

Once the rough location $(r_c,\phi_c)$ of the companions is known from the detection pipeline, our spectral extraction pipeline is composed of the following steps:
\begin{itemize}
\item[1.] Minimization of wavelength-to-wavelength cross talk: We partition the images in such a way that ensures that there is no companion signal in the search zones. For the wavelengths $\lambda_0 < \lambda_{Mid}$, where $\lambda_{Mid} = 1380$ nm is the central wavelength of the Project 1640 bandpass, we choose search zones azimuthally centered at $\phi_c$ over the radial interval $[r_c - N_{\delta} W,   r_c - N_{\delta} W +\Delta r]$ and only use reference images with $\lambda > \lambda_0 $.  For the wavelengths $\lambda_0 > \lambda_{Mid}$ we choose search zones azimuthally centered at $\phi_c$ over the radial interval   $[r_c + N_{\delta} W -\Delta r,   r_c + N_{\delta} W ]$ and {\em only use reference images with $\lambda < \lambda_0 $}. This ensures that there is no companion signal in the search zones but dramatically reduces the number of PSFs in the reference ensemble when only using slices from same exposure. We alleviate this issue by aggregating slices at all the relevant wavelengths from the full observing sequence in a larger  reference ensemble. We then proceed through KLIP reduction for these search zones over the range of parameters for which the companion has been previously detected. 
\item[2.] Calibration of the self-subtraction using forward modeling: Using un-occulted PSFs we apply the forward modeling methodology described in \citet{2012ApJ...755L..28S}. We use an un-occulted PSF at each wavelength as a template. For each reduction parameter this process produces single wavelength detection maps of width $9$ pixels around $(r_c,\phi_c)$. We then derive the spectro-photometry in each channel as value at the peak of an area of the detection map contained in a small circle of diameter $3$ pixels centered at $(r_c,\phi_c)$. The random spectro-photometric uncertainty is estimated as the scatter within the $63 \%$ confidence interval of the detection map around this peak. The rather large $9 \times 9$ spatial extent of the detection maps is chosen as a sanity check to rule out the cases of residual speckles contaminating the companion's signal.  This is necessary to rule out the pathological case for which the tail of a speckle located $\sim 2$ pixels away from the companion is considered to be companion flux by the matched filter.
\item[3.] Estimation of the systematic uncertainty due to the reduction: The previous steps yield a series of spectra over a large collection of reduction parameters $(N_{\delta}, \Delta r, \Delta \phi,K_{\rm KLIP})$. Direct inspection of the detection maps determine whether or not the companions have been detected for a given set of parameters. For a given sub-set of  $(N_{\delta}, \Delta r, \Delta \phi)$ where detection occurs there are three regimes of PCA truncation: $K_{\rm KLIP} < K_{\rm KLIP, Min}$ for which residual speckles in the neighborhood of the companion are still visible; $K_{\rm KLIP} > K_{\rm KLIP, Max}$ for which detection occurs but the morphology of the PSF has been significantly altered by the self-subtraction and thus cannot be calibrated by forward modeling; and $K_{\rm KLIP, Min} < K_{\rm KLIP} < K_{\rm KLIP, Max}$ which are of scientific interest. We  analyze our extracted spectra in conjunction with single wavelength detection maps in order to establish the subset of parameters over which each planet is detected. The spectra are then derived as the mean of this subset of spectra and their uncertainty due to the reduction is derived as the scatter of this ensemble---for HR~8799 b and c this subset is rather large ($\sim 100$ spectra) while is is smaller for HR~8799 d and e ($\sim 10$ spectra).
\item[3.] Spectral calibration: The resultant spectra from the previous steps are expressed in fractional units of un-occulted core intensity of the primary star (e.g units of contrast). We derive the spectrum of HR~8799 using the A5V template standard from the Pickles library \citep{1998PASP..110..863P} normalized at m$_{\rm H} = 5.28$, or M$_{\rm H} = 2.30$ \citep{csv03} and binned at the Project 1640 resolution. The final spectra are obtained  by multiplying the results from the previous step with this stellar spectrum. Note that this procedure for spectral calibration is slightly different than in earlier Project 1640 results presented in \citet{hob10} and \citet{rrb12}.  Those relied on the ratio of aperture photometry estimates of both the PSF cores and companion. As a sanity check we conducted an analysis analogous to \citet{hob10} and \citet{rrb12}: we used the relevant reduced images selected in step 2 and conducted step 3.  By projecting un-occulted PSFs on the relevant principal components and using aperture photometry to calibrate the self-subtraction, we computed the spectral calibration using aperture photometry of un-occulted images. The two yielded results that are consistent within our uncertainties. 
\end{itemize}

\begin{figure}[ht]
\center
\resizebox{0.4\hsize}{!}{\includegraphics[angle=0.]{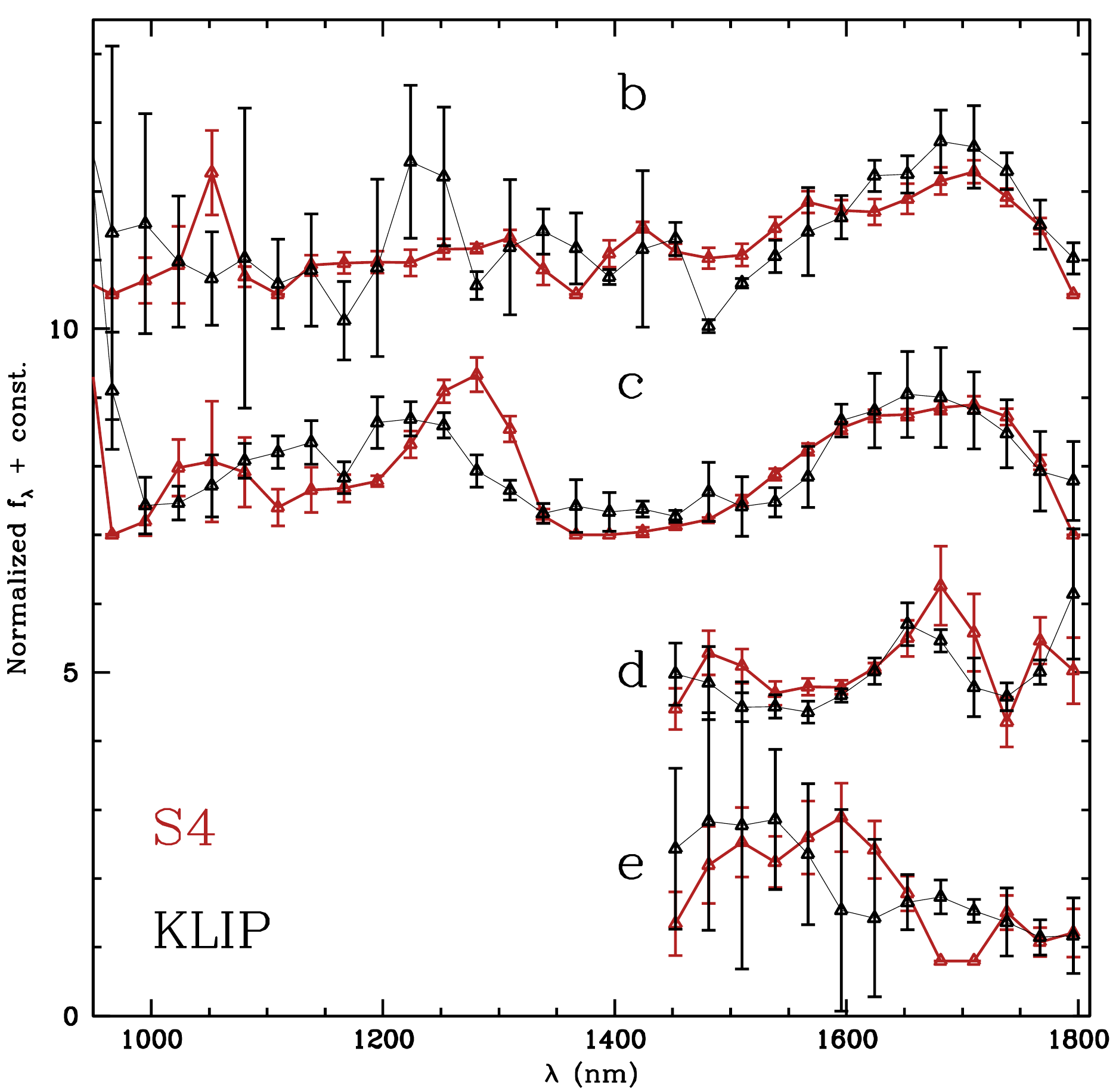}}
\caption{Comparison of spectra extracted with the KLIP algorithm and the S4 algorithm (in red).  Spectra are normalized to 1 and shifted by a constant for comparison.  Points that are weaker than a 2-$\sigma$ detection are excluded in the normalization and all points in $Y$ and $J$ bands are excluded for planets d and e.  Only a few of the 94 points plotted for each algorithm are discrepant by more than 2-$\sigma$, indicating good agreement between the two techniques}
  \label{klipvss4}
\end{figure}

\section{The S4 Speckle Suppression Algorithm}\label{s4sect}

\subsection{Detection}
The S4 algorithm for post-processing speckle suppression \citep{s4},
based on principal component analysis (PCA), has not yet been
published.  Thus we provide some details of it here.

S4 takes as input a 4D data block of dimension $ N_{\rm pixels_y}
\times N_{\rm pixels_x} \times N_{\lambda} \times N_{\rm Exposures}$
which has been pre-processed as follows: (i) application of a $2
\times 2$ median filter to each band/exposure to remove dead pixels;
(ii) spatial alignment
of all bands and exposures relative to one another and (iii) spatial
centering so that
the star lies precisely (within 0.1 pixel) at $(N_{\rm pixels_y}/2,N_{\rm pixels_x}/2)$.

\begin{figure*}[ht]
\center
\resizebox{1.\hsize}{!}{\includegraphics[angle=0.]{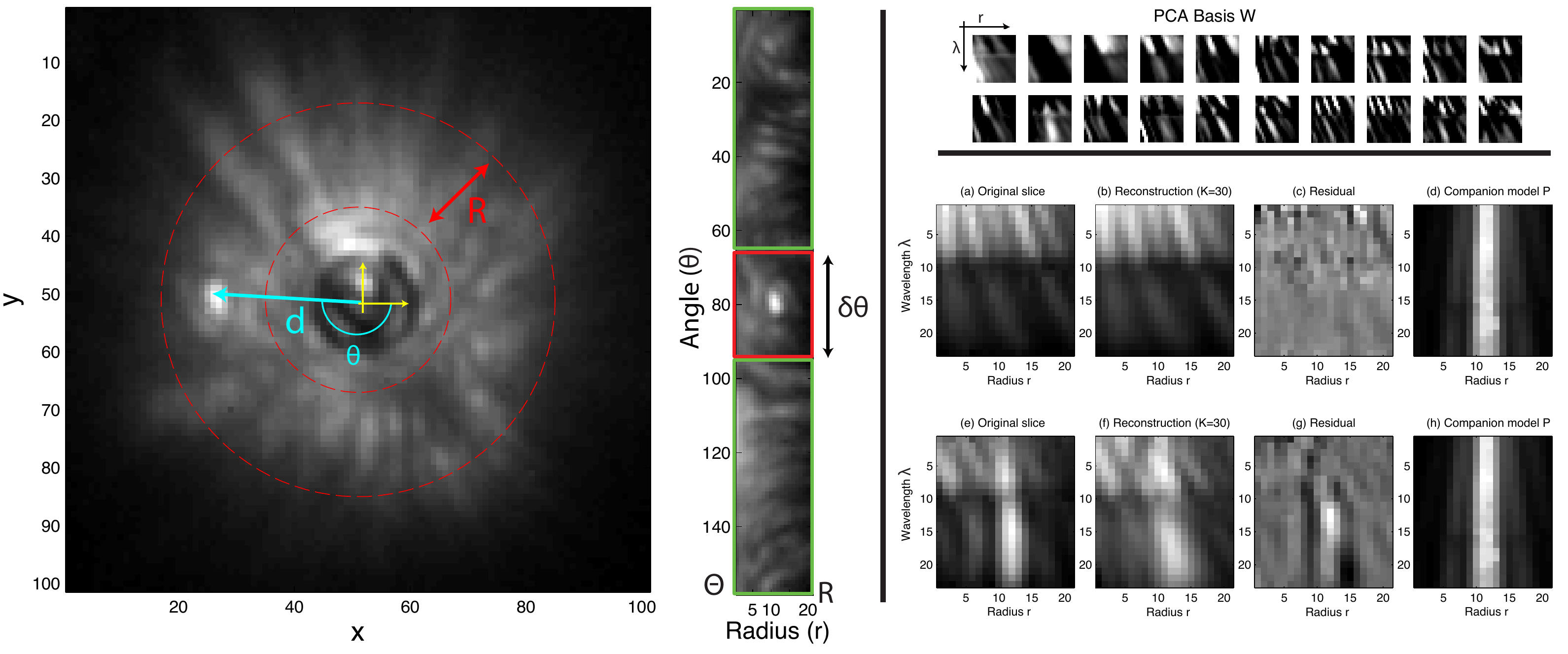}}
\caption{Schematic of the S4 Algorithm.  Left: Mean over wavelength $\lambda$ and $N$ exposures for an
  example star, FU Ori, whose companion is obvious, along with a polar representation of the annular
  region at radius $d$. Green and red regions shown training and test
  zones, respectively. Right top: The PCA basis $W$, computed from the
  training zone, showing the diagonal structure of the speckles in the
  joint radius-wavelength space. Right bottom: Reconstructions using
the PCA model with $K=30$ components.  (a): Original data
  slice, containing only speckles. (b): Reconstruction
  using PCA model with $k=30$. (c): Error
  residual. Note the lack of structure. (d): Companion model,
  which has low correlation with (c). Next row: (e):  Slice
  containing companion. (f): Reconstruction of PCA model. 
(g): The error residual shows clear structure
  associated with the companion, i.e.~the PCA speckle model cannot
  reconstruct the companion signal.  (h): Companion model,
  which has a high correlation with (g).}
  \label{s4fig}
\end{figure*}

In S4 the data at each spatial position is decomposed into a speckle
component and a companion component, whose sum reconstructs
the original data to the limit of Gaussian noise. To model the
speckle component effectively, which evolves radially with wavelength, this
decomposition is performed in a polar reference frame, as shown in
Fig.~\ref{s4fig} (left). To examine a location at radius $d$ and angle
$\theta$ from the center, an annular region of width $R$ at a radius $d$ from
is transformed into a region of size $\Theta \times R
\times N_{\lambda} \times N_{\rm Exposures}$, where $\Theta=2\pi d$ to
ensure the region is well-sampled. This region is divided into a $\emph
test$ zone around the location $\theta$ (of size $\delta
\theta$) and a $\emph training$ zone of all other angles, as shown in
red and green respectively in Fig.~\ref{s4fig} (left). We use the training
zone to build a model of the speckles that is then applied to the test
region, decomposing it into speckle and companion components.

We treat the speckles in both the training and test zones as being
independent over angle $\theta$ and exposure $n$, justified by limited
angular extent of the speckles and their variation due to atmospheric
turbulence and instrument flexure or temperature variations between exposures. We thus assume the structure of the
speckles to be confined to a joint radius-wavelength space (of
dimension $\lambda R$), as illustrated in
Fig.~\ref{s4fig} (right). This joint space is modeled using PCA, which
approximates each $\lambda R$-dimensional slice of the data at angle
$\theta$ cube $n$ as a linear combination of $K$ orthonormal basis
vectors $W=[w_1,\ldots,w_K]$ ($K$ being
a user-defined parameter). These basis vectors capture the majority of
the variance
in the radius-wavelength space and are computed by performing an
eigendecomposition of the covariance matrix built from the training
region, reshaped
into a matrix of $\lambda R$ dimensions by $(\Theta-\delta \theta)
N_{\rm Exposures}$ samples. We then use the PCA basis $W$ to infer the
speckle component of the test zone. $W$ is visualized in
Fig.~\ref{s4fig} (upper right).

In detection, we first use the PCA basis $W$ to fit the data in the
test zone and then correlate the residual error with a fixed companion
model $P$ (previously obtained by calibration with a point white light
source). The response for position $(\theta,d)$ is stored in a
correlation map. The intuition is that the PCA basis $W$ will effectively model
the speckles, but not the companion, thus the residual should only
contain the companion signal which will respond strongly to the
matched filter $P$ (see Fig.~\ref{s4fig}, lower right). The location
of the test zone is systematically
moved across all angles $\theta$ at a given radius, with the PCA
basis being recomputed at each location, because the training zone also
changes. The process is then repeated for a new radius $d$, until all
spatial locations in the visual field have been examined.

Finally, a normalization is performed on the correlation map that
compensates for the varition in flux with radius in the original
data. The resulting normalized map is then converted back to the Cartesian
coordinate frame and is the output of the detection algorithm.  (See
Fig.~\ref{detmap} and \ref{s4fig}(c) for examples.)

\subsection{Spectrum Extraction}

Promising peaks in the normalized detection map are then selected for
spectrum extraction. This follows the same overall modeling approach
as detection, except that the spectrum of the companion model $P$ is
no longer fixed
to be white. For a peak at location $(d,\theta)$, let the observed
data for exposure $n$ be $y_n$ (represented as $\lambda R$ dimensional
vector). We now must estimate
both the spectrum of the planet $m$ (an $N_\lambda$ dimensional vector)
and the $K$ dimensional PCA coefficients $z_n$ for exposure $n$ that
reconstruct $y_n$. Using a Gaussian noise model, this is equivalent to
minimizing the following convex objective:
\begin{equation}
\sum_{n=1}^{N_{\rm Exposures}} \| y_n - (Wz_n + Pm)\|^2_2
\end{equation}
where $Wz_n$ is the speckle component for exposure $n$ and $Pm$ is the
companion component (constant across all exposures). We impose a
non-negativity constraint on $m$, since negative spectra are not
physically plausible. Minimization is performed using standard
optimization software in Matlab. The spectra shown in
Fig.~\ref{S4spectra} are the resulting $m$ vectors for the 4 different
planet locations. An estimate of the uncertainty is obtained by measuring the
variance in $m$ over different settings of the model parameters,
$K$.  In the case of the data presented here, the number of principal components was varied from 50 to 500 with steps of 25 and the angular fitting size, $\Theta$, was 3, 5 and 7 pixels.  The error bars were calculated using principal components of 150, 175 and 200 and $\Theta$ of 5 and 7 pixels, settings which yielded the best results.  For October, due to significantly worse observing conditions, more principal components were needed for detection and spectrum extraction: 250, 300 and 350 with $\Theta$ of 5 and 7 pixels.

\bibliography{MasterBiblio}
\bibliographystyle{apj.bst}



\end{document}